\documentclass[fleqn,usenatbib]{mnras}

\usepackage{newtxtext,newtxmath}
\usepackage{multirow}

\usepackage[T1]{fontenc}

\DeclareRobustCommand{\VAN}[3]{#2}
\let\VANthebibliography\thebibliography
\def\thebibliography{\DeclareRobustCommand{\VAN}[3]{##3}\VANthebibliography}

\usepackage{graphicx}	
\usepackage{amsmath} 
\usepackage{subcaption}
\usepackage{caption}

\usepackage{adjustbox}  
\newcommand\orcid[1]{\href{http://orcid.org/#1}{\adjustbox{trim={-.15\width} {0\height} {-.15\width} {0\height},clip}{\includegraphics[height=10pt]{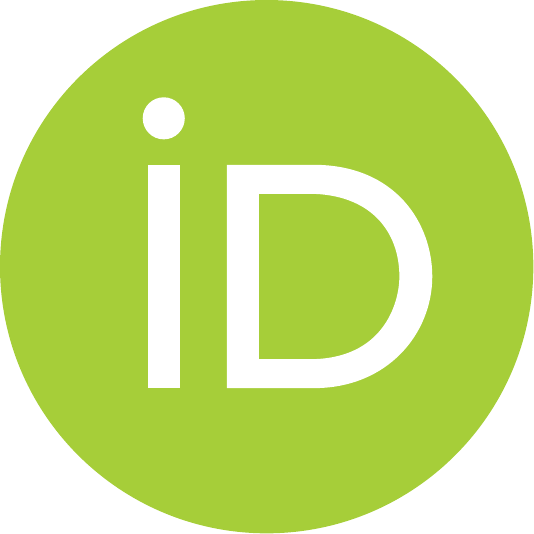}}}}

\title[Stellar masses of optically dark galaxies]{Stellar masses of optically dark galaxies: uncertainty introduced by the attenuation law and star-formation histories}

\author[Y. Lapasia et al.]{Yash Lapasia\orcid{0009-0009-5697-7965}$^{1}$\thanks{E-mail: yash.lapasia@gmail.com},
Sandro Tacchella\orcid{0000-0002-8224-4505}$^{1, 2}$\thanks{E-mail: st578@cam.ac.uk},
Francesco D'Eugenio\orcid{0000-0003-2388-8172}$^{1,2}$,
Dávid Puskás\orcid{0000-0001-8630-2031}$^{1,2}$,
\newauthor
Andrew J. Bunker\orcid{0000-0002-8651-9879}$^{3}$,
A. Lola Danhaive\orcid{0000-0002-9708-9958}$^{1,2}$,
Benjamin D. Johnson\orcid{0000-0002-9280-7594}$^{4}$,
Roberto Maiolino\orcid{0000-0002-4985-3819}$^{1,2}$,
\newauthor
Brant Robertson\orcid{0000-0002-4271-0364}$^{5}$,
Charlotte Simmonds\orcid{0000-0003-4770-7516}$^{6}$,
Irene Shivaei\orcid{0000-0003-4702-7561}$^{7}$,
Christina C. Williams\orcid{0000-0003-2919-7495}$^{8}$,
\newauthor
Christopher Willmer\orcid{0000-0001-9262-9997}$^{9}$,
Mengyuan Xiao\orcid{0000-0003-4102-794X}$^{10}$
\\
$^{1}$Kavli Institute for Cosmology, University of Cambridge, Madingley Road, Cambridge, CB3 0HA, UK\\
$^{2}$Cavendish Laboratory, University of Cambridge, 19 JJ Thomson Avenue, Cambridge, CB3 0HE, UK\\
$^{3}$Department of Physics, University of Oxford, Denys Wilkinson Building, Keble Road, Oxford OX1 3RH, UK\\
$^{4}$Center for Astrophysics $|$ Harvard \& Smithsonian, 60 Garden St., Cambridge MA 02138 USA\\ 
$^{5}$Department of Astronomy and Astrophysics, University of California, Santa Cruz, 1156 High Street, Santa Cruz, CA 95064, USA \\
$^{6}$Departamento de Astronomía, Universidad de Chile, Camino El Observatorio 1515, Las Condes, Santiago, Chile \\
$^{7}$Centro de Astrobiolog\'ia (CAB), CSIC-INTA, Ctra. de Ajalvir km 4, Torrej\'on de Ardoz, E-28850, Madrid, Spain \\
$^{8}$NSF National Optical-Infrared Astronomy Research Laboratory, 950 North Cherry Avenue, Tucson, AZ 85719, USA \\
$^{9}$Steward Observatory, University of Arizona, 933 North Cherry Avenue, Tucson AZ, 85721, USA \\
$^{10}$Department of Astronomy, University of Geneva, Chemin Pegasi 51, 1290 Versoix, Switzerland}

\pubyear{\the\year{}}

\begin{document}
\label{firstpage}
\pagerange{\pageref{firstpage}--\pageref{lastpage}}
\maketitle

\begin{abstract} 
Recent studies using \textit{JWST} have suggested that some high-redshift galaxies may be ultra-massive, thereby challenging standard models of early galaxy formation. We analyse the stellar masses using different modelling assumptions and in conjunction with new data of three galaxies (S1, S2 and S3), whose photometric and NIRCam/grism redshifts were consistent with $z>5$. These three ``optically dark'' galaxies were reported to host exceptionally high stellar masses ($M_{\star}\gtrsim10^{11}~M_{\odot}$) and star-formation rates ($\mathrm{SFR}\gtrsim600~M_{\odot}~\mathrm{yr}^{-1}$), implying extremely high star-formation efficiencies. Recent NIRSpec/IFU observations for S1 indicate a spectroscopic redshift of $z_{\rm spec}=3.2439\pm0.0002$, revising the previous redshift. Using the Bayesian spectral energy distribution (SED) modelling tool \texttt{Prospector}, we investigate the impact of key model assumptions on stellar mass estimates, such as the choice of star-formation history (SFH) priors (constant versus rising SFH base for the non-parametric SFH prior), the dust attenuation law, and the treatment of emission line fluxes. Our analysis yields revised stellar masses of $\log_{10}(M_{\star}/M_{\odot}) \approx 10.36^{+0.47}_{-0.32}, 10.95^{+0.11}_{-0.10}$ and $10.31^{+0.24}_{-0.19}$ for S1, S2, and S3, respectively. We find that adopting a rising SFH base prior results in lower inferred stellar masses compared to a constant SFH base prior. Additionally, we identify a significant degeneracy between the dust attenuation curve slope, the amount of dust attenuation, and stellar mass. Our results highlight various systematics in SED modelling due to SFH priors and dust attenuation that can influence stellar mass estimates. Nevertheless, even with these revised stellar mass estimates, two of the three galaxies remain among the most massive and actively star-forming systems at their respective redshifts, implying high star-formation efficiencies.
\end{abstract}

\begin{keywords}
galaxies: evolution -- galaxies: formation -- galaxies: high-redshift -- galaxies: star formation
\end{keywords}



\section{Introduction}

Unravelling the formation and evolution of galaxies across cosmic time remains one of the foremost challenges in modern astrophysics. In recent years, our understanding of the early Universe -- particularly during the Epoch of Reionization (EoR) -- has been transformed by ultra-deep observations from cutting-edge facilities such as the James Webb Space Telescope (\textit{JWST}) and the Atacama Large Millimeter/submillimeter Array (\textit{ALMA}). These telescopes have opened new observational windows onto the distant Universe, revealing populations of galaxies that were previously inaccessible. Among the most intriguing discoveries are the so-called ``optically dark'' galaxies: extremely red systems that elude detection in traditional optical bands but emerge clearly at infrared and submillimetre wavelengths. Specifically, these galaxies are characterized by non-detections shortward of $\sim1.5$–$2,\mu$m (i.e., undetected in HST bands to $H > 27$ AB mag) and robust detections at longer wavelengths \citep[e.g.,][]{dey99, wang09, simpson14, franco18, wang19_nature, talia21, gomez-guijarro22}. Many of these systems were not even detected at near-infrared wavelengths \citep{williams19}, so little was known about their stellar population properties. However, a key aspect of many of these sources with submillimetre only detections is that the spectral energy distribution (SED) limits basically mandated that they are at redshifts $z>2$. \textit{JWST}, delivering imaging and spectroscopy at wavelengths of $3$–$5\mu$m, now enables us for the first time to trace the rest-frame optical of these optically dark galaxies, providing critical insight into the obscured side of early galaxy formation \citep{barrufet23, barrufet26, gomez-guijarro23, rodighiero23, williams24}.

Accurate estimates of galaxy stellar masses are essential for tracing galaxy evolution, as they inform key properties such as the underlying stellar populations, star-formation histories (SFHs), and the timing of mass assembly. In this context, optically dark galaxies -- despite their faintness at rest-frame UV and optical wavelengths -- are of particular importance. These systems could potentially harbour substantial stellar mass and star formation, i.e., they may represent a significant fraction of early galaxy growth. If omitted, their lack might introduce biases in the derivation of both the galaxy stellar mass function and the cosmic star formation history, particularly at high redshifts. Recent analyses have highlighted the non-negligible contribution of dust-obscured galaxies to the total cosmic star formation rate density (SFRD). For instance, pre-JWST studies estimated that optically dark galaxies contribute $\sim3-6\%$ of the SFRD at $z\sim3-4$ \citep{casey14, novak17, gruppioni20, fudamoto20, zavala21}. Recent JWST studies indicate that the level of dust obscuration at $z>4$ in massive galaxies can be quite high and points to a substantial fraction of the SFRD (up to $30-40\%$) at these early epochs between $3<z<6$ being obscured \citep{barrufet23, barrufet26, gottumukkala24, williams24, cheng25, martis25, sun25_aspire}. These obscured systems appear to dominate the high-mass end ($M_{\star} \gtrsim 10^{10}~M_\odot$) of the star-forming population during this epoch \citep{gomez-guijarro23, xiao23, wang25_n}. Constraining their stellar masses and star-formation rates (SFRs) is therefore critical to obtaining a complete census of galaxy growth in the early Universe. 

\textit{JWST} provides an unprecedented opportunity to study the rest-frame optical emission of optically dark galaxies, granting access to the stellar continuum features that are essential for constraining their stellar masses. However, inferring accurate stellar masses for these heavily obscured systems remains a challenge. The first critical requirement is a reliable redshift measurement to precisely determine the galaxy's distance. Even with a secure redshift, deriving the stellar mass requires accurate knowledge of the mass-to-light ratio ($M_{\star}/L$), which depends on the age, dust attenuation, and metallicity of the stellar population. A key diagnostic feature in this context is the Balmer/4000 \AA~ break. However, interpreting this feature relies on the assumption that the observed continuum is dominated by stellar light. In high-$z$ galaxies, nebular emission -- especially strong emission lines at rest-frame wavelengths $\lambda_{\rm rest} > 4000$ \AA~ -- can significantly boost broadband fluxes, potentially mimicking or enhancing an apparent Balmer or 4000 \AA~ break \citep{schaerer09, stark13, tacchella22_highz, endsley24}. Furthermore, a new, JWST-discovered class of active galactic nuclei (AGN), so called ``Little Red Dots'' (LRDs), can show Balmer breaks that are non-stellar origin, caused by very dense gas absorbing the AGN continuum \citep{inayoshi25_break, ji25_break, deugenio26_blackthunder, naidu26}. Misinterpretation of this rest-optical emission as stellar can dramatically increase the inferred stellar masses of photometric candidates \citep{williams24}. Finally, given the potentially severe dust attenuation in optically dark, high-$z$ galaxies, the choice of attenuation law directly influences the shape of the SED \citep[e.g.,][]{maek18, markov23}. Specifically, one can easily hide one order of magnitude of the stellar mass with flatter attenuation \citep{lo-faro17, williams19}. While these uncertainties in properties pose formidable challenges to characterizing massive red galaxies, the importance of this population motivates the detailed study of methods to disentangle these systematics.

In a recent study, \citet{xiao24} use the data from the \textit{JWST} FRESCO survey \citep{oesch23} to constrain the photometry of a subset of 36 massive dust-obscured galaxies at a redshift of $z=5-9$. They find three ``ultra-massive'' galaxies (S1, S2 and S3) at redshifts $z \approx 5-6$ exhibiting stellar masses far exceeding expectations with $M_{\star}\gtrsim10^{11}~M_{\odot}$. Importantly, these three sources are spatially extended (see Fig.~\ref{fig:three_images}), different from most other ultra-massive galaxy candidates, which usually are unresolved. The extended morphology together with the strong cold dust emission implies that an AGN contribution is unlikely, making them among a small sample of contamination-free ultra-massive candidates \citep{xiao25}. \citet{xiao24} used photometry to measure stellar masses for S1, S2 and S3, inferring baryon-to-stellar conversion efficiency in these galaxies is about 50\%, which is two to three times higher than the highest efficiencies observed at lower redshifts. \par

In this paper, we reassess stellar masses for these three optically dark galaxies of \citet{xiao24}. Recent NIRSpec/IFU observations indicate a new spectroscopic redshift for S1 ($z_{\rm spec}=3.2439 \pm 0.0002$ instead of $z=5.579$ as previously reported, see \citealt{xiao26}). We include broader photometric coverage, including more \textit{JWST} imaging data, H$\alpha$ emission line fluxes, and FIR photometry from ALMA/NOEMA (Section ~\ref{sec:sample}). In Section~\ref{sec:sed_modelling} we introduce our SED modelling setup: we use \texttt{Prospector} \citep{johnson19}, allowing for a flexible dust attenuation law and exploring a range of SFH priors. We present our results in Section~\ref{sec:sed_results}, discuss their implications in Section~\ref{sec:discussion} and conclude in Section~\ref{sec:conclusions}. Throughout this paper we assume a flat $\Lambda \text{CDM}$ cosmology.

\section{Sample and observational data}
\label{sec:sample}

In this section, we describe the galaxy sample analysed in this work and summarise the observational data used to constrain their physical properties. We first introduce the three optically dark galaxies selected from the \textit{JWST}/FRESCO survey and review the assumptions underlying their originally reported redshifts and stellar masses. We then detail the multi-wavelength data set employed here, including deep \textit{JWST} imaging, slitless grism spectroscopy, and (sub-)millimetre measurements, which together form the basis for our analysis.

\subsection{Three optically dark galaxies}

In this work, we focus on the three most massive galaxies identified in \citet{xiao24}: S1, S2 (also known as GN10), and S3 (see Table~\ref{table:galaxy property}). These galaxies were selected from the FRESCO survey \citep{oesch23, covelo-paz25}, which provides imaging in three NIRCam filters (F182M, F210M, and F444W) and F444W grism spectroscopy, covering approximately 124 arcmin$^2$ across the GOODS-S and GOODS-N fields. The selection targeted galaxies with red colours, defined by $\mathrm{F182M} - \mathrm{F444W} > 1.5$ mag -- a threshold consistent with colour cuts used to identify optically faint or ``dark'' galaxies \citep[e.g.,][]{gomez-guijarro23} -- and exhibiting emission lines suitable for redshift determination. All selected sources were required to have a $>5\sigma$ detection in F444W, and for objects undetected in F182M, a $2\sigma$ upper limit was used to place a lower limit on the colour.

\citet{xiao24} identified 36 dusty star-forming galaxies (DSFGs) at $z=4.5-9.1$ and estimated their stellar masses, dust attenuation, and SFRs by fitting \textit{JWST}/FRESCO and HST photometry with the BAGPIPES code \citep{carnall18}. Their analysis adopted a constant SFH, the \citet{calzetti00} dust attenuation law, and fixed redshifts based on the grism-derived measurements. Nebular emission was not explicitly modelled in the SED fits; instead, the contribution of strong emission lines was removed from the F444W flux using grism-based measurements. From this parent sample, the three most massive systems consistent with redshifts $z = 5-6$ were selected: S1, S2, and S3 -- all of which exhibit extremely red colours, with $\mathrm{F182M} - \mathrm{F444W} > 3.5$ mag.

These sources were not newly discovered by \textit{JWST}; they were previously detected in ground-based SCUBA-2 observations \citep{dannerbauer08, cowie18}. In particular, S2 (GN10) has long-standing multi-wavelength coverage and redshift estimates predating \textit{JWST} \citep[and references therein]{riechers20}. S1 has also been observed with ALMA \citep{gomez-guijarro22} and \textit{JWST}.

Fitting the three-band NIRCam photometry from FRESCO, \citet{xiao24} reported that S1, S2, and S3 host extremely massive stellar populations ($\log_{10}(M_\star/M_\odot) \gtrsim 11.0$) and exhibit substantial dust attenuation ($A_V > 3$ mag). While S1 is detected in both F182M and F210M, S2 is undetected in these bands (with only upper limits), and S3 is detected only in F210M. There is no evidence for a significant AGN contribution to the rest-frame optical light based on emission-line properties, source morphology, and ancillary multi-wavelength data. If located at $z \sim 5.8$, the stellar masses of these galaxies would imply host halo masses of $\log_{10}(M_{\text{halo}}/M_\odot) > 12.5$. \citet{xiao24} concluded that such high stellar-to-halo mass ratios would require star-formation efficiencies 2-3 times greater than the maximum observed at low redshift ($\varepsilon_{\text{max,obs}} \simeq 0.2$), presenting a significant challenge to current galaxy formation models.

\begin{table}
\centering
\begin{tabular}{|c|c|c|c|c|}
\hline
Object & RA (1) & Dec (2) & $z_{\text{grism}}$ (3) & $z$ (4) \\
\hline
S1 & 03:32:28.91 & -27:44:31.53 & 5.579 & 3.244 \\
S2  & 12:36:33.42 & +62:14:08.57 & 5.306 & 5.306 \\
S3  & 12:36:56.56 & +62:12:07.37 & 5.179 & 5.179 \\
\hline
\end{tabular}
\caption{Galaxy properties of the three galaxies. (1) Right ascension (J2000); (2) Declination (J2000); (3) NIRCam/grism redshifts presented in \citet{xiao24}; (4) Redshifts adopted in this work. For S2 and S3, we adopt the same spectroscopic redshifts as in \citet[][see also \citealt{riechers20, sun24_filaments}]{xiao24}, while for S1 we use the updated spectroscopic redshift obtained from NIRSpec/IFU observations of $z_{\rm spec}=3.2439 \pm 0.0002$ \citep{xiao26}.}
\label{table:galaxy property}
\end{table}

\begin{figure*}
    \centering
    \begin{subfigure}[b]{0.3\textwidth}
        \includegraphics[width=\textwidth]{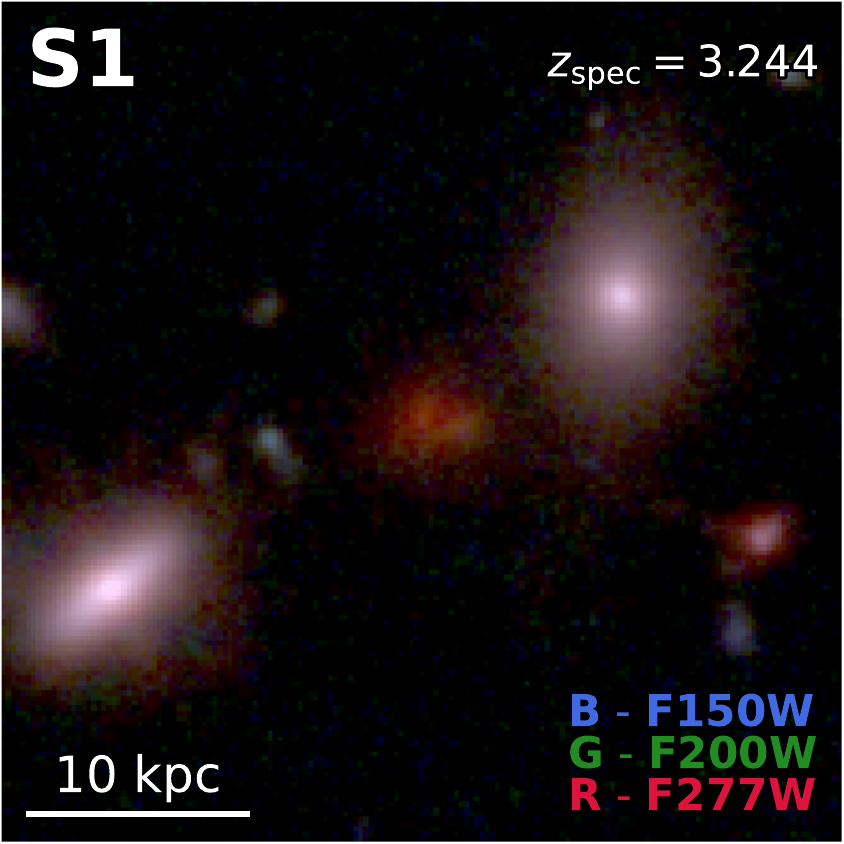}
        \label{fig:image1}
    \end{subfigure}
    \hfill
    \begin{subfigure}[b]{0.3\textwidth}
        \includegraphics[width=\textwidth]{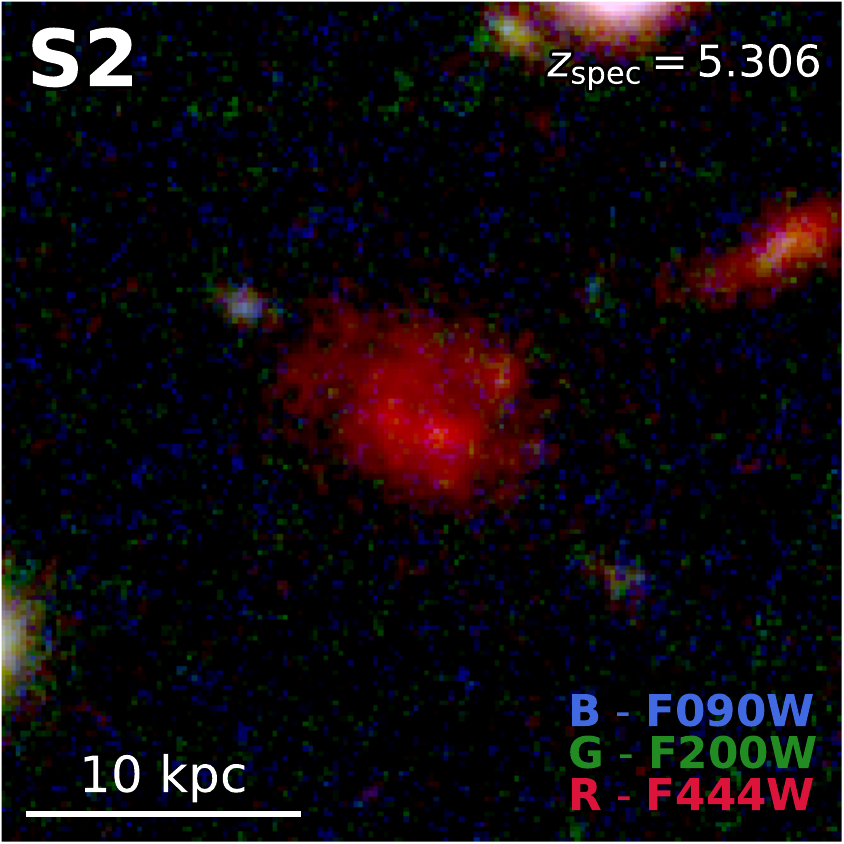}
        \label{fig:image2}
    \end{subfigure}
    \hfill
    \begin{subfigure}[b]{0.3\textwidth}
        \includegraphics[width=\textwidth]{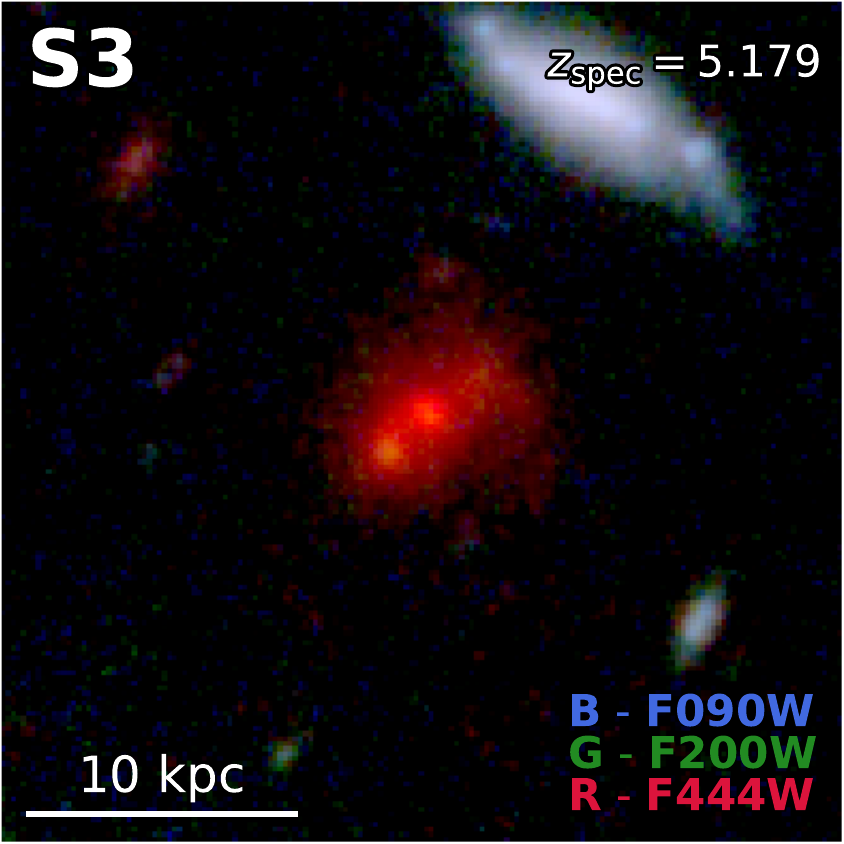}
        \label{fig:image3}
    \end{subfigure}
    \caption{NIRCam RGB images of S1, S2 and S3 from left to right. The red/green/blue colours correspond to F277W/F200W/F150W for S1 and F444W/F200W/F090W for S2 and S3. The scale bars indicate a projected physical distance of 10~kpc, computed at the respective galaxy redshift using the angular diameter distance. S1 has a faint second component in the south-west (toward the lower right). S2 and S3 are more extended and show multiple components, potentially indicative of ongoing or recent merger activity.}
    \label{fig:three_images}
\end{figure*}

\begin{table*}
\centering
\begin{tabular}{|c|c|c|c|}
\hline
\textbf{Filter} & \textbf{S1 (nJy)} & \textbf{S2 (nJy)} & \textbf{S3 (nJy)} \\ 
\hline
F090W      & $8.7 \pm 13.65$                   & $70.9 \pm 53.3$         & $38.0 \pm 40.3$   \\ 
F115W      & $-3.5 \pm 11.8$                   & $0.3  \pm 35.4$         & $21.9 \pm 26.5$   \\ 
F150W      & $89.3 \pm 15.4$                   & $30.6  \pm 44.4$        & $54.1 \pm 29.0$   \\ 
F182M      & $278.7 \pm 19.4$                  & $80.1 \pm 39.4$         & $10.7 \pm 76.2$   \\ 
F200W      & $227.0 \pm 11.3$                  & $68.0 \pm 36.1$         & $188.6 \pm 27.5$  \\ 
F210M      & $282.6 \pm 26.8$                  & $31.4\pm 50.9$          & $147.4 \pm 112.8$ \\ 
F277W(A)   & $701.6 \pm 35.1$                  & NA                       & $712.4 \pm 35.6$  \\
F277W(B)   & $601.2 \pm 30.0$                  & $302.3 \pm 21.5$        & NA                 \\
F335M(A)   & $916.9 \pm 45.8$                  & NA                       & $712.1 \pm 35.6$  \\ 
F335M(B)   & NA                                 & $478.7 \pm 28.8$        & NA                 \\ 
F356W(A)   & $1131.2 \pm 56.6$                 & NA                       & $788.2 \pm 39.4$  \\
F356W(B)   & $1081.6 \pm 54.1$                 & $595.5 \pm 29.8$        & NA                 \\ 
F410M(A)   & $1713.3 \pm 85.7$                 & NA                       & $1633.0\pm 81.7$   \\
F410M(B)   & NA                                 & $1282 \pm 62.4$         & NA                  \\ 
F444W(A)   & $2159.3 \pm 107.9$                & $1290 \pm 64.5$         & $1531.5 \pm 76.6$   \\ 
F444W(B)   & $2022.5 \pm 101.1$                & $1249 \pm 62.4$         & $1317.4 \pm 77.3$   \\  
H$\alpha$ flux (erg$\text{s}^{-1}\text{cm}^{-2}$) & NA                        & $(1.17 \pm 0.1) \times10^{-17}$ & $(1.495 \pm 0.1) \times10^{-17}$ \\
ALMA 1.1mm      & $(950 \pm 120)\times10^3$    & NA                         & NA                 \\
NOEMA 1.0mm     & NA                            & $(955 \pm 73)\times10^4$  & NA                 \\
NOEMA 1.2mm     & NA                            & $(525 \pm 60)\times10^4$  & NA                 \\
NOEMA 2.2mm     & NA                            & $(28 \pm 17)\times10^4$  & NA                 \\
NOEMA 2.7mm     & NA                            & $(14.8 \pm 3.2)\times10^4$  & NA                 \\
\hline
\end{tabular}
\caption{Observed photometry for S1, S2, and S3, expressed in nJy. We use the Kron-convolved photometry. The measurements include NIRCam imaging from FRESCO (F182M, F210M, F444W) and JADES (F090W to F444W) surveys, with long-wavelength channel filters separated into modules A and B due to differing response curves. H$\alpha$ line fluxes (in erg$\text{s}^{-1}\text{cm}^{-2}$) are included for S2 and S3 from FRESCO grism spectroscopy, along with far-infrared fluxes from ALMA (S1) and NOEMA (S2) at $1-3$ mm. Filters and observations not available are marked as NA (Not Available).}
\label{tab:fluxes}
\end{table*}

\subsection{Observational data}

The photometric data used in our analysis are listed in Table~\ref{tab:fluxes}. In addition to the FRESCO data \citep{oesch23}, we incorporate observations from the \textit{JWST} Advanced Deep Extragalactic Survey \citep[JADES;][]{eisenstein26}. Specifically, in addition to the F182M, F210M, and F444W bands provided by FRESCO, we include photometry in the F090W, F115W, F150W, F200W, F277W, F335M, F356W, and F410M filters and increased depth in F444W from JADES\footnote{\url{https://archive.stsci.edu/hlsp/jades}} \citep{rieke23, robertson24, deugenio25_DR3}. For the long-wavelength channel, we distinguish between the A and B modules, as they exhibit different throughput curves. Details on source detection and photometric extraction are provided in \citet{rieke23} and \citet{robertson24}. For S2 and S3, we further include H$\alpha$ line fluxes measured from the FRESCO grism spectroscopy \citep{xiao24, covelo-paz25, herard-demanche25}.

Composite images of all 3 galaxies using \textit{JWST}/NIRCam F277W, F200W and F150W (for S1) and F444W, F200W and F090W (for S2 and S3) filters are shown in Fig.~\ref{fig:three_images}. All three galaxies show some irregularities. S1 consists of two components, with a faint clump in the southwest (towards the lower right), which might be a satellite (see discussion below). S2 and S3 are extended and consist of several components. These features might indicate ongoing merger activity. 

While the rest-frame UV, optical, and NIR data from \textit{JWST} provide critical constraints on the stellar populations, the far-infrared (FIR) emission offers a powerful probe of the dust-reprocessed light. This is essential for energy balance modelling (see Section~\ref{sec:sed_modelling}), as the FIR directly quantifies the starlight absorbed by dust and re-emitted at longer wavelengths, enabling more accurate determination of intrinsic stellar population properties, including star formation rates and dust attenuation. S1 is detected in the GOODS-ALMA 1.1 mm survey in GOODS-S \citep{franco18, gomez-guijarro22}, while S2 benefits from extensive NOEMA coverage at $1-3$ mm \citep{dannerbauer08, riechers20}.

\begin{figure}
    \centering
    \includegraphics[width=\linewidth]{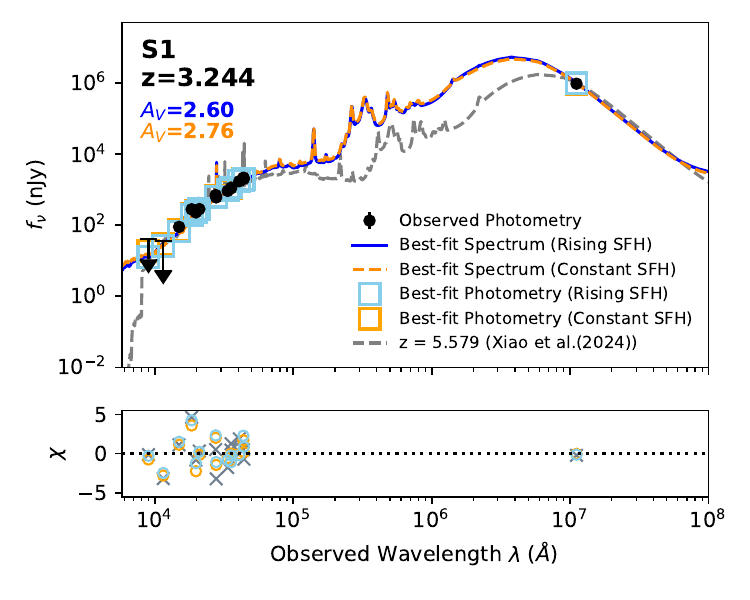}
    \caption{\textit{Top:} Best-fitting SED for S1 with the redshift fixed to the spectroscopic value $z_{\rm spec}=3.244$ (blue solid line for RSFH prior and dashed orange line for CSFH prior) and, for comparison, to the redshift adopted by \citet{xiao24}, $z_{\text{Xiao}} = 5.579$ (grey dashed line). Black points show the observed photometric data, including the ALMA 1.1\,mm measurement (see Table~\ref{tab:fluxes}). \textit{Bottom:} The normalized residuals ($\chi$), i.e., photometric residuals between the observed and model-predicted fluxes, normalised by the observational uncertainties. Blue and orange circles correspond to the fit using the new spectroscopic redshift $z_{\rm spec}=3.244$ for the RSFH and CSFH prior, respectively, while grey crosses show the residuals obtained when fixing the redshift to $z_{\text{Xiao}} = 5.579$. The photometry favours the new spectroscopic redshift, yielding a total $\chi^2 = 42.5$ and 37.7 for the RSFH and CSFH prior, respectively, compared to $\chi^2 = 53.7$ for the fit adopting the redshift from \citet{xiao24}.}
    \label{fig:S1_zspec}
\end{figure}

\begin{table*}
\centering
\begin{tabular}{|c|c|c|}
\hline
\textbf{Parameter} (1) & \textbf{Description} (2) & \textbf{Prior} (3) \\
\hline
$z_{\text{red}}$ & fixed to the spectroscopic redshift & $z_{\text{spec}}$ \\
$\log Z/Z_\odot$ & Stellar metallicity & $\mathcal{G}(0.0,0.5,-1.0,0.0)$ \\
$\log M_{\star}/M_\odot$ & Total stellar mass formed & $\mathcal{U}(6,12)$ \\
SFH & Flexible SFH Prior with 8 time bins & Constant SFH Prior and Rising SFH Prior \\
$n_{\text{dust}}$ & Power-law modifier to shape of the \citet{calzetti00} dust attenuation curve of the diffuse dust & $\mathcal{U}(-1.0,0.4)$ \\
$\tau_2$ & Diffuse dust optical depth & $\mathcal{G}(0.3,1,0.0,6.0)$ \\
$\tau_1$ & Birth cloud optical depth, fitted for ratio $\tau_1/\tau_2$  & $\mathcal{G}(1.0,0.3,0.0,2.0)$\\
$\log Z_{\text{gas}}/Z_\odot$ & Gas-phase metallicity & $\mathcal{U}(-2.0,0.5)$ \\
$\log U$ & Ionization parameter for the nebular emission & $\mathcal{U}(-4,-1)$\\
$\gamma_{\rm e}$ & Fraction of dust exposed to very strong radiation fields & $\mathcal{U}(0.001,0.15)$ \\
$q_{\rm PAH}$ & Fraction of dust in the form of PAHs & $\mathcal{U}(0.5,7.0)$\\
$U_{\rm min}$ & Minimum intensity of the radiation field heating most of the dust & $\mathcal{U}(0.1,25)$ \\
\hline
\end{tabular}
\caption{Parameters and their priors used in our fiducial Model. (1) Name of the Parameter (2) Description of the parameter (3) Parameter prior probability distribution; $\mathcal{U}(a,b)$ is a uniform distribution between $a$ and $b$; $\mathcal{G}(\mu,\sigma,a,b)$ is a clipped Gaussian distribution with mean $\mu$ and dispersion $\sigma$ between $a$ and $b$.}
\label{table:priors and parameters}
\end{table*}

\subsection{Redshifts}
\label{subsec:redshift}

\citet{xiao24} inferred redshifts from \textit{JWST}/NIRCam grism spectroscopy obtained as part of the FRESCO survey. For S2 and S3, the grism observations clearly reveal both H$\alpha$ and [S II] emission lines, enabling robust redshift measurements. For S2 (also known as GN10), the derived redshift is $z = 5.3064 \pm 0.0005$, consistent with $z = 5.303 \pm 0.001$ from VLA CO observations \citep{riechers20} and $z = 5.30 \pm 0.01$ from an independent analysis of the FRESCO data \citep{sun24_filaments}. For S3, the redshift is measured to be $z = 5.1793 \pm 0.0004$ \citep{xiao24}.

In contrast, the redshift of S1 was more uncertain and formed the subject of detailed discussion in \citet{xiao24}. The FRESCO slitless grism data reveal only a single emission line, which is slightly offset from the centroid of the continuum emission. This offset could arise from spatial variations in line equivalent width across the galaxy or from contamination by a nearby source along the dispersion direction, as S1 lies in a crowded region. As a result, the possibility of line misidentification or blending cannot be robustly excluded based on the grism data alone. To resolve this ambiguity and obtain a secure redshift measurement, \citet{xiao26} therefore analyse dedicated \textit{JWST}/NIRSpec IFU observations of S1. Our independent inspection of the NIRSpec/IFU spectrum confirms the multiple emission-line identifications and the revised redshift. In Appendix~\ref{app:pagamma}, we describe our measurement of the Pa$\gamma$ flux. After correcting for attenuation, the Pa$\gamma$ luminosity implies an SFR of approximately $71\pm16\,M_\odot\,\mathrm{yr}^{-1}$., consistent within the Prospector-derived $\mathrm{SFR}_{50}$ values of $104^{+58}_{-33}$ and $89^{+26}_{-27}\,M_\odot\,\mathrm{yr}^{-1}$ for the RSFH and CSFH priors, respectively.

\begin{figure*}
    \centering
    \includegraphics[width=1.\textwidth]{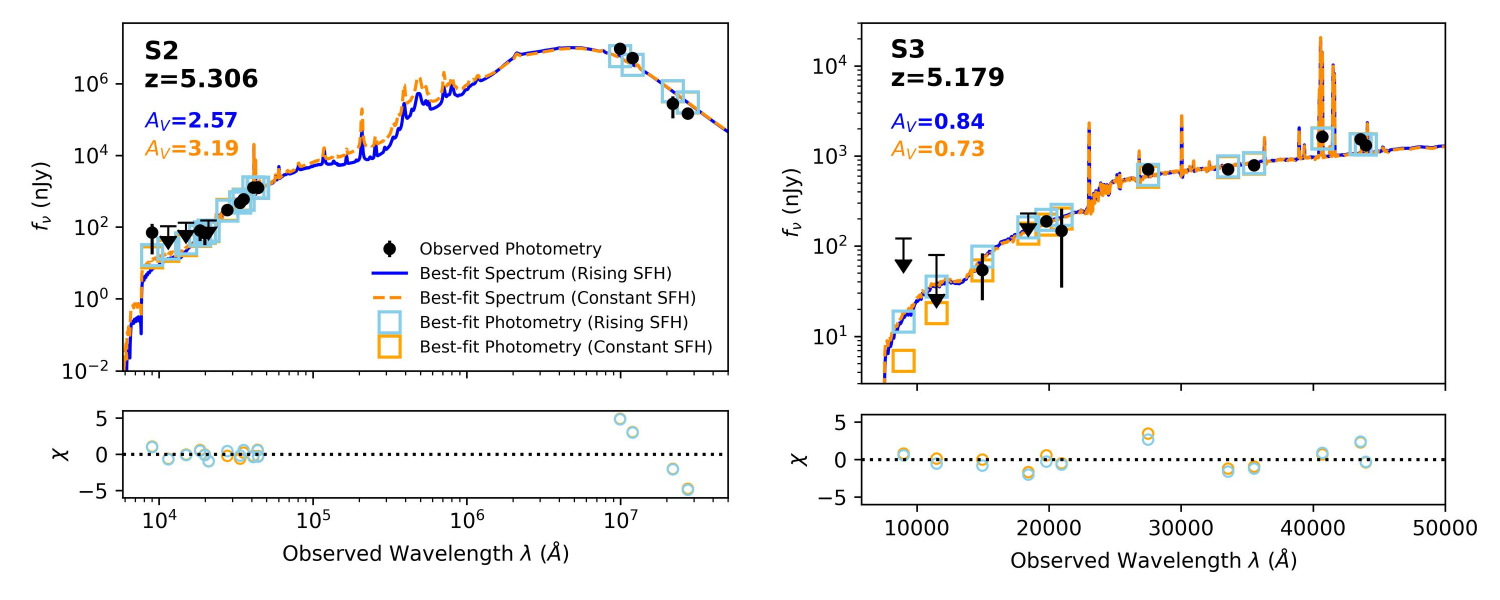}
    \caption{\textit{Top:} SED fits for galaxies S2 (left panel) and S3 (right panel). Black points show photometric measurements for several filters shown in Table \ref{tab:fluxes}, with black error bars indicating observational uncertainties. NOEMA data for S2 is also included in the fit. The blue solid and orange dashed lines show the best-fit spectra from \texttt{Prospector} using rising and constant SFH priors, respectively. Blue and orange squares mark the corresponding model-predicted photometry. \textit{Bottom:} The normalized residuals ($\chi$), normalized by the observational uncertainties, are generally distributed around zero for both galaxies, indicating a good overall fit for both the rising and constant SFH models. However, for both S2 and S3, the residuals from the rising SFH fits (blue circles) appear slightly more tightly clustered and symmetric around $\chi \approx 0$ compared to those from the constant SFH fits (orange circles), suggesting that the rising SFH prior provides a marginally better fit to the observed photometry. This is supported by the total $\chi^2$ values: for S2, $\chi^2_{\text{rising}} = 64.7$ and $\chi^2_{\text{constant}} = 64.8$; and for S3, $\chi^2_{\text{rising}} = 23.5$ and $\chi^2_{\text{constant}} = 24.2$.}
    \label{fig:sed}
\end{figure*}

\begin{table*}
    \centering
    \begin{tabular}{|c|cc|cc|cc|cc|cc|}
        \hline
        & \multicolumn{2}{c|}{$\log(M_\star/M_\odot)$} & \multicolumn{2}{c|}{$\text{SFR}_{50}~[M_\odot/\text{yr}]$} & \multicolumn{2}{c|}{$t_{50}$ [Gyr]} & \multicolumn{2}{c|}{$\tau_2$} & \multicolumn{2}{c|}{$n_{\text{dust}}$} \\
        \hline
        Object & RSFH & CSFH & RSFH & CSFH & RSFH & CSFH & RSFH & CSFH & RSFH & CSFH \\
        \hline
        S1 & $10.36^{+0.47}_{-0.32}$ & $10.74^{+0.22}_{-0.36}$ & $104^{+58}_{-33}$ & $89^{+26}_{-27}$ & $0.12^{+0.31}_{-0.09}$ & $0.32^{+0.32}_{-0.17}$ & $2.41^{+0.33}_{-0.44}$ & $2.56^{+0.40}_{-0.38}$ & $0.07^{+0.19}_{-0.16}$ & $0.23^{+0.10}_{-0.15}$ \\[0.15cm]

        S2 & $10.95^{+0.11}_{-0.10}$ & $11.22^{+0.07}_{-0.12}$ & $536^{+167}_{-81}$ & $579^{+170}_{-110}$ & $0.12^{+0.14}_{-0.07}$ & $0.25^{+0.08}_{-0.16}$ & $2.38^{+0.25}_{-0.11}$ & $2.95^{+0.13}_{-0.14}$ & $0.01^{+0.11}_{-0.08}$ & $0.17^{+0.06}_{-0.11}$ \\[0.15cm]
        S3        & $10.31^{+0.24}_{-0.19}$ & $10.51^{+0.18}_{-0.33}$ & $67^{+65}_{-34}$ & $38^{+50}_{-19}$ & $0.15^{+0.19}_{-0.10}$ & $0.43^{+0.25}_{-0.36}$ & $0.78^{+0.26}_{-0.17}$ & $0.68^{+0.24}_{-0.17}$ & $-0.52^{+0.30}_{-0.26}$ & $-0.47^{+0.25}_{-0.35}$ \\
        \hline
    \end{tabular}
    \caption{Summary of fiducial runs for all three galaxies, presenting key parameters: stellar mass ($M_\star$), SFR averaged over the past 50 Myr (SFR$_{50}$), mass-weighted stellar age ($t_{50}$), optical depth of diffuse dust attenuation ($\tau_2$), and dust attenuation curve index ($n_{\text{dust}}$). Results are shown for both the constant SFH base prior (CSFH) and the rising SFH base prior (RSFH).}
    \label{table:result}
\end{table*}

Specifically, galaxy S1 was observed with \textit{JWST}/NIRSpec \citep{ferruit22} in IFU mode \citep{boker22}, as part of Cycle~3 GO program~5572 (PI: Xiao), using the high-resolution G395H grating in combination with the F290LP filter. This configuration provides continuous wavelength coverage over $2.87\text{--}5.55\,\mu$m at a spectral resolving power of $R\approx2700$. As discussed in \citealt{xiao26}, both \text{[S\,{\sc{iii}}]$\lambda$9531} and \text{He\,{\sc{i}}$\lambda 1.083\mu$m} show broad and double-peaked spectral profiles. By inspecting the NIRCam imaging (Fig.~\ref{fig:three_images}), we identify a sub-structure with bluer colour in the SW, which could contribute to the complex line profiles. The detected emission lines unambiguously place S1 at a spectroscopic redshift of $z_{\rm spec}=3.2439 \pm 0.0002$ \citep[see][]{xiao26}, which is significantly lower than the published redshift from \citet{xiao24} of $z = 5.5793 \pm 0.0006$. Besides the unambiguous redshift provided by four detected emission lines (the two aforementioned lines, plus lower-significance \text{[S\,{\sc{iii}}]$\lambda$9069} and Pa\,$\gamma$), the aperture spectrum of S1 contains no line signal near 4.31 $\mu$m (the observed wavelength of H$\alpha$ at $z=5.5793$); the 4.31-$\mu$m emission line reported in \citet[][with flux $F\approx6\times10^{-18}~\mathrm{erg\,s^{-1}\,cm^{-2}}$]{xiao24} would be easily detected by NIRSpec in $\approx 6.5$~hours, which reaches a ten times lower sensitivity than FRESCO \citep{oesch23}. We adopt the NIRSpec/IFU spectroscopic redshift from \citet{xiao26} as the fiducial redshift for S1 throughout this work.

In Fig.~\ref{fig:S1_zspec}, we compare SED fits with \texttt{Prospector} obtained by fixing the redshift to this new spectroscopic value and, for comparison, to the redshift adopted by \citet{xiao24}, $z_{\text{Xiao}} = 5.5793$. In both cases, the same photometric data and SED modelling setup are used (see next section). The residuals across the SED demonstrate that the fit at the new spectroscopic redshift provides a slightly better description of the data, with a total $\chi^2 = 42.5$ and 37.7 (for the RSFH and CSFH prior, respectively), compared to $\chi^2 = 53.7$ when fixing the redshift to $z_{\text{Xiao}}$. This highlights that photometric redshifts remain a challenging measurement for extremely dusty sources, even with JWST.

\section{SED modelling with Prospector}
\label{sec:sed_modelling}

\texttt{Prospector} is a state-of-the-art SED modelling code that employs a fully Bayesian framework to infer galaxy properties from photometry and spectroscopy \citep{johnson19,johnson21}. It features modelling of stellar populations, nebular emission, and dust attenuation, and supports flexible, non-parametric SFHs \citep{leja19_nonparm}. A strength of \texttt{Prospector} is its enforcement of energy balance between UV–optical absorption and IR re-emission, enabling physically consistent treatment of both stellar and dust emission components. Its modular design and robust uncertainty quantification make it particularly powerful for interpreting deep \textit{JWST} observations, where dust, emission lines, and variable SFHs are critical. The use of non-parametric SFHs is especially advantageous for DSFGs at high redshift, as it allows the data -- including the FIR constraints -- to reveal complex, bursty, or rapidly rising star formation episodes without imposing restrictive functional forms.

\begin{figure*}
    \centering
    \includegraphics[width=\linewidth]{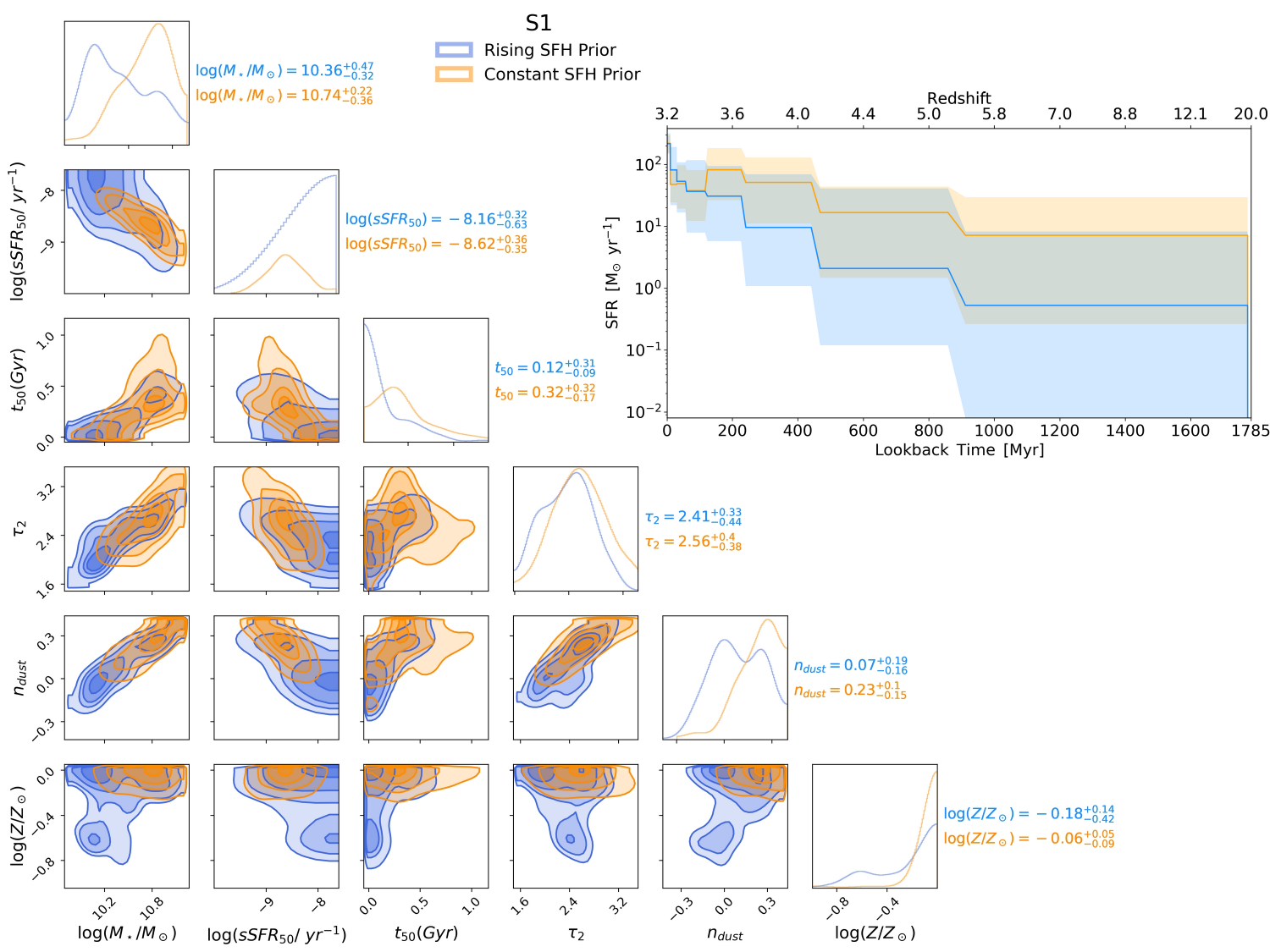}
    \caption{Posterior distribution of key stellar population parameters: total stellar mass formed ($M_{\star}$), specific star formation rate averaged over 50 Myr (sSFR$_{50}$), mass-weighted stellar age ($t_{50}$), optical depth of the diffuse dust attenuation ($\tau_2$), dust attenuation curve index ($n_{\rm dust}$), and stellar metallicity ($Z_{\star}$). Results are shown for two SFH base priors: rising SFH (RSFH; blue) and constant SFH (CSFH; orange). Median values for each parameter under both priors are indicated. The top-right panel shows the resulting SFH, illustrating that the CSFH prior prefers a higher SFR at earlier times than the RSFH prior, which gives rise to a higher stellar mass and a lower sSFR.}
    \label{fig:corner_s1}
\end{figure*}

\begin{figure*}
    \centering
    \includegraphics[width=\linewidth]{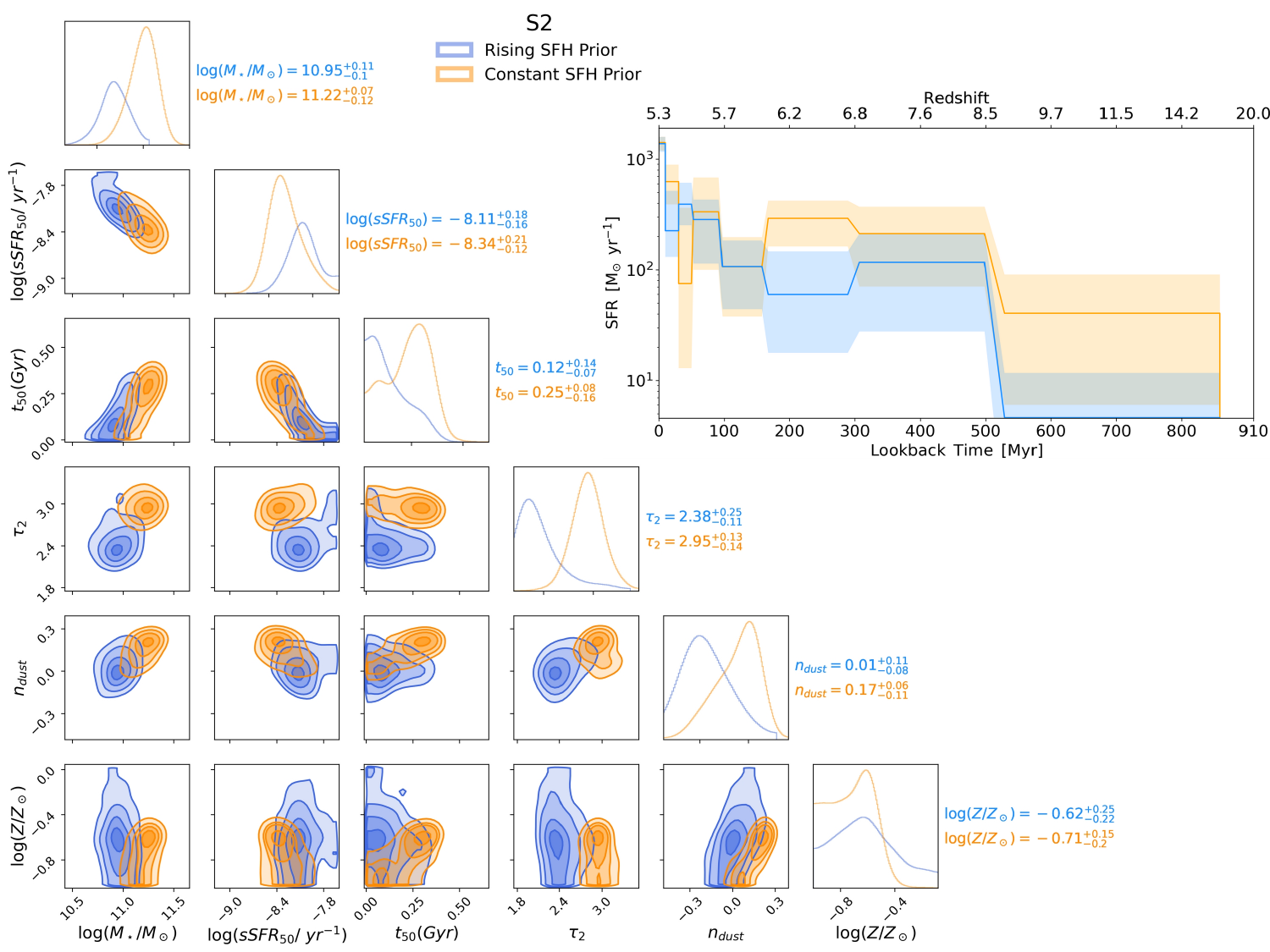}
    \caption{Same as Fig.~\ref{fig:corner_s1}, but for S2. The CSFH prior converges on a slightly higher stellar mass solutions, with a lower sSFR and older stellar age than the RSFH prior.}
    \label{fig:corner_s2}
\end{figure*}

\begin{figure*}
    \centering
    \includegraphics[width=\linewidth]{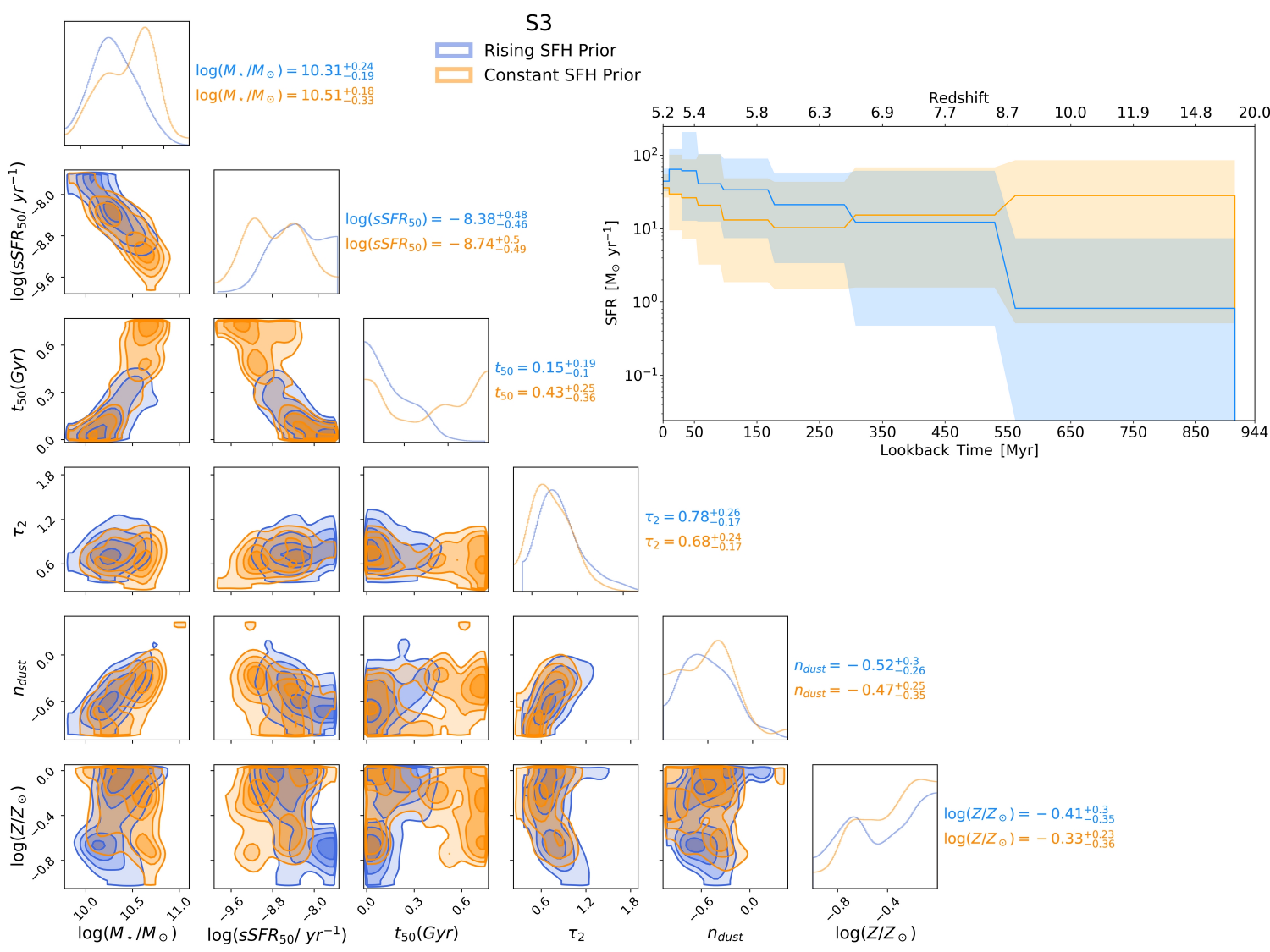}
    \caption{Same as Fig.~\ref{fig:corner_s1}, but with S3. Both SFH priors return multimodal posterior distributions, with the CSFH prior leading to a higher SFR at $z>9$, which gives rise to a higher stellar mass, lower sSFR and older stellar age.}
    \label{fig:corner_s3}
\end{figure*}

We follow largely the \texttt{Prospector} model described in \citet{tacchella22_highz} and \citet{tacchella23_metal}. Specifically, we jointly fit the photometry and H$\alpha$ emission line flux (when available, namely for S2 and S3) with \texttt{Prospector}. Table \ref{table:priors and parameters} provides a detailed overview of the free parameters used in the fiducial model, along with their respective priors. In brief, we utilize the MESA Isochrones \& Stellar Tracks \citep[MIST;][]{choi16, dotter16} and the \texttt{MILES} spectral library \citep{sanchez-blazquez06}. We assume the \citet{chabrier03} initial mass function (IMF) with mass cutoffs of 0.1 and 100 M$_{\odot}$. The prior of the stellar metallicity $\log(Z_{\star}/Z_{\odot})$ is a normal centred on 0.0 (i.e., solar metallicity) and a width of 0.5, clipped at $-1.0$ (i.e., 10\% solar metallicity) and $0.0$. 

We use a two-component dust model from \citet{charlot00} and \citet{conroy09a}, which accounts for the increased effect of dust on young stars and nebular emission ($<10$ Myr) and couples the dust attenuation of the diffuse component to the UV dust bump at 2175 \AA~ \citep{kriek13}. The dust attenuation is modelled by three parameters: the birth cloud optical depth $\tau_1$, the diffuse optical depth $\tau_2$, and a power-law modifier to shape of the \citet{calzetti00} dust attenuation curve of the diffuse dust $n_{\rm dust}$. The priors are given in Table \ref{table:priors and parameters}. This flexibility in the dust attenuation modelling is crucial to assess possible degeneracies between the property of the stellar populations (such as stellar age and metallicity) and dust attenuation, in particular for dusty galaxies such as the ones considered here. For the nebular model, which includes both emission lines and nebular continuum, we use the \texttt{Cloudy} model grids \citep[v13.03;][]{ferland13} as presented in \citet{byler17}, self-consistently propagating the ionising flux from the stars. This nebular model has two free parameters: the ionisation parameter $U$ and the gas-phase metallicity $Z_{\rm gas}$. 

We assume the standard energy balance, i.e., all the energy attenuated by dust is re-emitted in the IR. Thanks to this assumption, the FIR photometry from ALMA and NOEMA delivers additional constraints on the total amount of dust attenuation and on the dust-free stellar SED. We use the \citet{draine07} dust emission templates to describe the shape of the IR SED, which are based on the silicate-graphite-PAH model of interstellar dust. These templates have three free parameters controlling the shape of the IR SED, which we allow to vary in our modelling: ($i$) $\gamma_{\rm e}$, which represents the fraction of dust heated by intense radiation fields; ($ii$) $q_{\rm PAH}$, the fraction of dust mass in the form of PAHs; and ($iii$) $U_{\rm min}$, which measures the minimum radiation field intensity (in units of the local interstellar radiation field) responsible for heating the dust. 

Leveraging the flexibility of \texttt{Prospector}, we incorporate two distinct SFH priors. Both SFH priors are non-parametric in shape (i.e., do not assume a parametric form of the SFR as a function of time), but assume the SFH consists of 8 bins in lookback time. The first bins are $0-10$ Myr and $10-30$ Myr, while the remaining are logarithmically spaced back to redshift $z=20$. \texttt{Prospector} fits for the log-ratio of the SFR in these bins, meaning that there are 7 free parameters plus the total stellar mass, which is the integral over the SFH. The prior for the log-ratio of the SFH is a Student's t-distribution with $\sigma = 0.5$, allowing for bursty star formation \citep{tacchella22_highz, danhaive25_disc}. The difference between the two SFH priors assumed in this work relies on the base of the SFH. The first prior is the standard ``continuity'' prior \citep{leja19_nonparm}, where the base SFH is constant, i.e., the Student-t distribution of the log-ratios of adjacent bins is centred on 0. We call this the constant SFH (CSFH) prior in this work. The second prior used in this work (RSFH) is a rising prior \citep{turner25}, which assumes that galaxies' SFR follows  $\mathrm{SFR}(z) \propto \exp[-4/5\cdot(z-z_{\rm obs})] \cdot (1 + z)^{5/2}$, i.e., tracking dark matter halo assembly. This leads to a preference of rising SFHs at early cosmic times ($z>3$). This is particularly important in not overestimating stellar masses of galaxies if they undergo a recent burst of star formation, since older stellar populations will be outshone by younger ones, which introduces significant uncertainty in the stellar ages and stellar masses \citep[e.g.,][]{whitler23_sfh, tacchella22_highz, tacchella23_metal}.

In summary, our \texttt{Prospector} model has 17 parameters (Table~\ref{table:priors and parameters}). We fix the redshift to the spectroscopic redshift listed in Table~\ref{table:galaxy property}. We present results for both the rising (RSFH) and constant (CSFH) base of the non-parametric SFH prior. 

The best-fit SEDs are shown in Fig.~\ref{fig:S1_zspec} for S1, while S2 and S3 are shown in Fig.~\ref{fig:sed} on the left and right, respectively. For both S2 and S3, the fits from the two SFH priors are broadly consistent, with minor differences visible in the residuals ($\chi$) shown below each SED. In both cases, the rising SFH prior yields a marginally better fit, with total $\chi^2$ values of 64.7 and 23.5 for S2 and S3, respectively, compared to 64.8 and 24.2 for the constant SFH prior.

\section{Stellar population properties}
\label{sec:sed_results}

In this section, we present the stellar population properties inferred from our fiducial \texttt{Prospector} modelling of the three optically dark galaxies. We focus on the robustness of the derived stellar masses and SFHs, and quantify how these depend on key modelling choices such as the assumed SFH prior and dust attenuation curve. We then place our revised stellar mass estimates in the context of previous results and discuss the implications for the interpretation of these extreme systems.

\subsection{Stellar masses, SFH and dust attenuation}
\label{subsec:stellar_mass_dust_law}

We first focus on key stellar population parameters, including stellar masses, SFHs and dust attenuation. To assess the impact of assumptions about SFHs on the inferred physical properties, we compare the posterior distributions obtained using two different SFH priors: a rising (RSFH) and a constant (CSFH) SFH base prior. As mentioned above, this base prior describes the base of non-parametric SFH, around which the SFR is allowed to vary. This comparison allows us to quantify the sensitivity of derived parameters such as stellar mass, dust attenuation, metallicity, and stellar age to the choice of prior. \par

For S1 (Fig.~\ref{fig:corner_s1}), the inferred physical parameters exhibit a strong dependence on the assumed SFH prior, but the posteriors generally overlap. The stellar mass increases from a lower value inferred under the rising SFH prior ($\log(M_{\star}/M_{\odot})=10.36_{-0.32}^{+0.47}$) to a significantly higher value under the constant SFH prior ($\log(M_{\star}/M_{\odot})=10.74_{-0.36}^{+0.22}$). The stellar mass and consequently the sSFR show strong degeneracies with nearly all other parameters. Similar degeneracies are evident for the dust attenuation curve index ($n_{\text{dust}}$) and the optical depth of the diffuse dust attenuation ($\tau_2$). Interestingly, the RSFH prior gives rise to a more multimodal posterior distribution than the CSFH. This could indicate a stronger degeneracy between the different parameters or it could reflect multiple stellar populations. For example, it is possible that a recently formed population of young stars has a lower metallicity and a grayer attenuation curve than the stars that formed more continuous over a longer time. Despite these differences, both priors consistently recover a recent starburst. Both priors indicate a rising SFH (top right panel in Fig.~\ref{fig:corner_s1}) with a $SFR_{50}$ of $\approx 100~M_\odot~\mathrm{yr}^{-1}$, but this trend is stronger for the RSFH prior as expected. This leads to a lower stellar mass, since the SFR at early cosmic times is consistently lower for the RSFH than the CSFH prior. \par

For S2 (Fig.~\ref{fig:corner_s2}), the inferred posterior distributions of the parameters are overall more constrained for both SFH priors (less multimodality). The stellar mass increases from $\log(M_{\star}/M_\odot)=10.95_{-0.1}^{+0.11}$ under the rising SFH prior to $11.22_{-0.12}^{+0.07}$ with the constant SFH prior, accompanied by corresponding shifts in $t_{50}$. Both priors display similar degeneracy structures, particularly among stellar mass, $\mathrm{sSFR}_{50}$, and $n_{\text{dust}}$. Both priors suggest a vigorous ongoing starburst, with $SFR_{50}$ of approximately $\approx536-579~M_\odot\mathrm{yr}^{-1}$. \par

For S3 (Fig.~\ref{fig:corner_s3}), the posteriors show multimodality for both SFH priors. The stellar mass increases slightly from $\log(M_{\star}/M_\odot)=10.31_{-0.19}^{+0.24}$ (RSFH prior) to $10.51_{-0.33}^{+0.18}$ (CSFH prior). Both priors lead to a rather strong degeneracy between stellar mass, $n_{\text{dust}}$, and $t_{50}$. While both priors lead to an overall increasing SFH, the SFR is significantly larger at $z>9$ for the CSFH prior than for the RSFH prior, which explains the stellar mass differences between the two. \par

Our fiducial SED fits presented in this section incorporate the full photometric data, including \textit{JWST} NIR, FIR (ALMA/NOEMA), and H$\alpha$ line fluxes when available. To assess the role of FIR data, we compare fits with and without these long-wavelength constraints in Appendix~\ref{app:stellar_mass_FIR}. For S1, the inclusion of ALMA flux does not significantly change the posterior distributions for all parameters, including stellar mass. On the other hand, S2 shows significantly tighter posteriors when NOEMA data are included, highlighting the importance of FIR measurements in constraining physical parameters. 

\subsection{Comparison to previous works}
\label{subsec:stellar_mass_comparison}

We briefly compare our stellar masses and dust attenuation values to the previous work of \citet{xiao24}. We adopt our fiducial, rising SFH base prior for this comparison. For S1, we get a stellar mass of $\log_{10}(M_{\star}/M_{\odot}) = 10.36^{+0.47}_{-0.32}$ at $z=3.2439$, while \citet{xiao24} found $\log_{10}(M_{\star}/M_{\odot}) \approx 11.34^{+0.11}_{-0.13}$ at $z=5.579$. If we assume the same redshift as \citet{xiao24}, i.e., $z=5.579$, we obtain a stellar mass of $\log_{10}(M_{\star}/M_{\odot}) \approx 10.10^{+0.10}_{-0.12}$, which makes the discrepancy even larger ($>1$ dex). The reason for this is two-fold. Firstly, we adopt a rising SFH prior, which -- as shown in the previous section -- prefers lower stellar masses and younger stellar ages. Secondly, we assume a flexible attenuation curve, while \citet{xiao24} fixes theirs, which is commonly adopted for heavily dust obscured sources in the absence of direct constraints on the shape of the attenuation curve. The impact of the attenuation curve on the stellar masses is discussed further in Section~\ref{subsec:ML_dust}.

For S2 (GN10), our stellar mass of $\log(M_{\star}/M_\odot) = 10.95^{+0.11}_{-0.10}$ is consistent with the one from \citet{xiao24} of $\log(M_{\star}/M_\odot) = 11.17^{+0.21}_{-0.21}$. \citet{riechers20} modelled GN10's [$C_{\text{II}}$] emission as a rotating disk and derived a dynamical mass of $4.5^{+2.1}_{-1.5} \times 10^{10}\,M_\odot$ within 3.66 kpc and $8.6^{+3.6}_{-2.8} \times 10^{10}\,M_\odot$ within 5.49 kpc. They also reported a stellar mass of $1.2 \pm 0.1 \times 10^{11}\,M_\odot$, in between the estimate from \citet{xiao24} and ours. These values exceed the dynamical mass budget, especially if the stellar and gas components are similarly extended, indicating a potential overestimate of the stellar mass or limitations in the dynamical modelling. This could indeed be the case since we find indication for a disturbed morphology, which could be merger induced (Fig.~\ref{fig:three_images}). 

For S3, our estimate of $\log(M_{\star}/M_\odot) = 10.31^{+0.24}_{-0.19}$ is significantly lower (by about $2\sigma$) than the $\log(M_{\star}/M_\odot) = 11.01^{+0.16}_{-0.15}$ value from \citet{xiao24}, which at least in part can be explained by the rising SFH prior. 

Since we showed in the previous section that the photometric data cannot fully break the degeneracy between stellar mass and dust attenuation, we now also compare the dust attenuation values between our work and the one from \citet{xiao24}. Importantly, while \citet{xiao24} uses a fixed Calzetti dust attenuation law, we fit for a two component dust model with a flexible attenuation curve for the diffuse component. The dust attenuation values in the $V$ band ($A_V$) we infer depend moderately on the assumed SFH prior. Under a constant SFH prior, our estimates for S1 and S2 ($A_V$ = 2.35 and 3.19) are broadly consistent with those reported by \citet{xiao24}, though systematically lower than their values of 3.2 and 3.4. Adopting a rising SFH prior shifts these two sources to $A_V\approx2.6$, slightly increasing S1 and noticeably reducing S2. S3 shows a larger discrepancy with \citet{xiao24} regardless of SFH choice: our values remain low ($A_V$ = 0.73-0.84) at both SFH priors compared to $A_V$ = 3.4 reported by \citet{xiao24}. However, S3 is the only source for which no FIR or submillimetre constraint is included in our SED modelling. Its dust attenuation is therefore constrained entirely by the rest-frame UV--optical photometry, without an independent energy-balance constraint on the absorbed luminosity. Within our adopted model, the posterior is relatively well constrained, with $A_V=0.84^{+0.26}_{-0.17}$ mag for the RSFH prior and $A_V=0.73^{+0.24}_{-0.17}$ mag for the CSFH prior. Nevertheless, these statistical uncertainties are conditional on the assumed SFH and attenuation model and do not capture the full systematic uncertainty. The discrepancy with \citet{xiao24} may reflect differences in the photometric coverage and SED-fitting methodology, particularly their adoption of a constant SFH and a fixed Calzetti attenuation law. The inferred $A_V$ for S3 should therefore be interpreted with greater caution than those of S1 and S2. \par

\section{Discussion}
\label{sec:discussion}

\begin{figure}
    \centering
    \includegraphics[width=\linewidth]{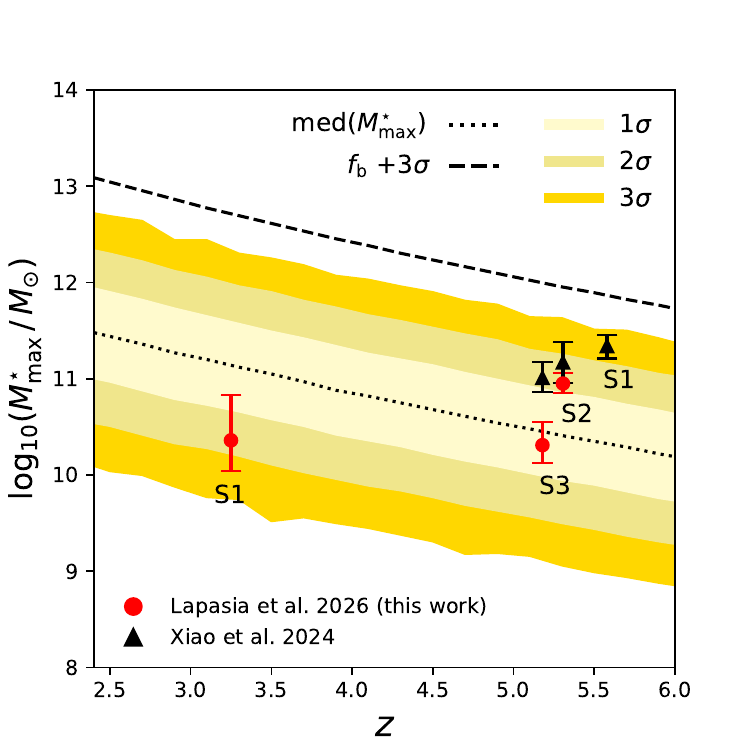}
    \caption{Maximum stellar mass expected in the survey volume of FRESCO, covering an area of 124 arcmin$^{2}$, as estimated with EVS confidence intervals. Stellar mass estimates for galaxies S1, S2, and S3 from this work (red circles) and from \citet{xiao24} (black triangles) are shown with their respective uncertainties. The dashed line represents the $3\sigma$ upper limit, assuming a stellar mass fraction of unity. All stellar mass estimates from \citet{xiao24} lie between the $2\sigma$ and $3\sigma$ limits, whereas the estimates for S2 and S3 from this work fall within $<2\sigma$. The stellar mass estimate for S1 lies significantly below the median maximum stellar mass constraint since we adopt the spectroscopic redshift $z_{\rm spec}=3.244$. Overall, while S2 and S3 are massive for their redshifts and consistent with high integrated star-formation efficiencies, these galaxies are consistent with expectations from galaxy formation models in the context of $\Lambda$CDM cosmology.}
    \label{fig:mass_z}
\end{figure}

\subsection{Implications for star-formation efficiency}

After revisiting the stellar properties and redshifts of these three optically dark galaxies, we now assess the star-formation efficiency of these galaxies and the implications for galaxy formation models in the context of the $\Lambda$CDM framework. In order to estimate the expectation of a galaxy with a certain stellar mass in a given cosmic volume, we use extreme value statistics (EVS) to generate confidence regions in the mass-redshift plane for the most extreme mass haloes and galaxies, following \citet{lovell23_evs}. Fig.~\ref{fig:mass_z} shows the EVS confidence intervals for the FRESCO survey area of 124 arcmin$^2$. The dotted and dashed lines show two different models of how to populate dark matter halos with galaxies. The dotted line assumes a truncated log-normal distribution for the stellar mass fraction with $\mu = e^{-2} \approx 0.135$. This is motivated from numerical simulations such as EAGLE \citep{crain15, schaye15}, FLARES \citep{lovell21, vijayan21_phot}, or THESAN \citet{kannan22_thesan}, and empirical models \citep{behroozi13a, tacchella18}. The dashed line shows the $3\sigma$ upper limit for a model that assumes all baryons are converted into stars (with a baryon fraction of $f_{\rm b} = 0.16$). Galaxies on this line would challenge the expectations of galaxy formation models in the $\Lambda$CDM context.

The revised redshift and stellar mass of S1 alter its interpretation compared to previous studies. At the lower redshift, we infer a stellar mass of $\log(M_\star/M_\odot) \approx 10.36$. We find that S1 exhibits substantial dust attenuation ($\tau_2 \approx 2.4$) and a $SFR_{50}$ of $\approx100~M_\odot$/yr. At cosmic noon ($z\approx1-3$), heavily dust-obscured star formation is typically associated with more massive ($M_{\star}>10^{10}~M_\odot$) galaxies \citep{whitaker17, shivaei24}. S1 demonstrates that even moderate-mass systems can reach extreme obscured fractions, perhaps due to compact starbursts or local ISM geometry. The galaxy is indeed slightly elongated and could be a dusty, edge-on disk galaxy \citep{nelson23_ufo, gibson24}.

\begin{figure*}
    \centering
    \includegraphics[width=\textwidth]{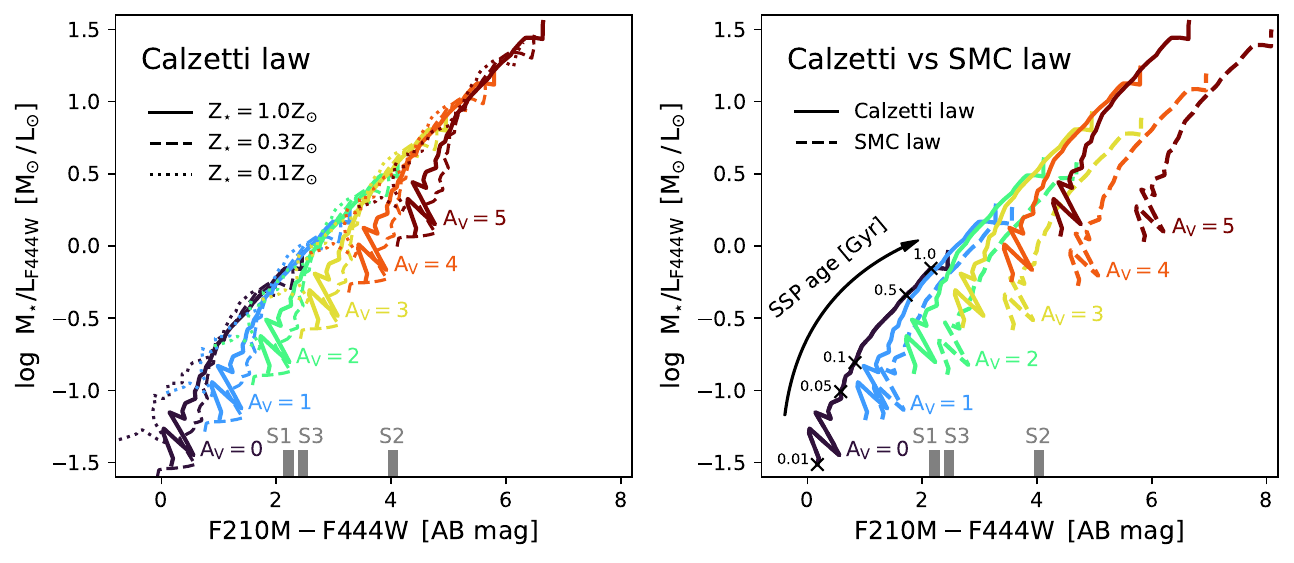}
    \caption{Stellar mass-to-light ratio ($\log\mathrm{M_{\star}/L_{\rm F444W}}$) as a function of the F210M$-$F444W colour at $z=5.5$. \textit{Left panel:} The dotted, dashed and solid lines show stellar population models assuming a Simple Stellar Population (SSP) and a stellar metallicity of $\mathrm{Z}_{\star}=0.1$, 0.3 and 1.0 Z$_{\odot}$, respectively. Along each line, the SSP age increases from 0.01 Gyr to 1.5 Gyr. The colour coding corresponds to the amount of dust attenuation A$_{\rm V}$, assuming a fixed \citet{calzetti00} law. The grey bars at the bottom indicate the F210M$-$F444W colours of S1, S2 and S3. \textit{Right panel:} The solid and dashed lines indicate the solar metallicity SSP models assuming the \citet{calzetti00} and SMC \citep{gordon03} law, respectively. The line colours represent different values of A$_{\rm V}$ as in the left panel. At older ages (SSP age of $>100$ Myr), the variations in age, metallicity and amount of dust attenuation collapse onto a narrow relation between $\mathrm{M_{\star}/L_{\rm F444W}}$ and F210M$-$F444W colour. However, this relation is different for different attenuation laws: the steeper SMC law leads to a lower $\mathrm{M_{\star}/L_{\rm F444W}}$ at fixed F210M$-$F444W colour.}
    \label{fig:ml_col}
\end{figure*}

As shown in Fig.~\ref{fig:mass_z}, the revised stellar mass estimates for S2 and S3 fall within the range predicted by the EVS confidence intervals, indicating no significant tension with theoretical expectations. The earlier estimates from \citet{xiao24} lie closer to the upper bounds of the EVS confidence intervals, suggesting a potential challenge to the model. Nevertheless, in both analyses, the stellar masses of S2 and S3 imply a high star-formation efficiency. High star-formation efficiencies are expected in overdense regions of the early Universe, such as proto-clusters \citep{chiang13, chiang17, lim24}. S2 (GN10) and S3 reside within such a dense environment at $z = 5.17-5.30$, which also hosts the luminous DSFG HDF850.1 \citep{calvi23, herard-demanche25, sun24_filaments}. This large-scale structure contains over 100 H$\alpha$-emitting galaxies identified by \textit{JWST}/FRESCO slitless spectroscopy. The disturbed morphologies of S2 and S3 point to ongoing or recent merger activity, reinforcing the view that dense environments enhance galaxy–galaxy interactions and trigger intense star formation. In this context, their large stellar masses may be at least partly attributable to mergers, which can supply substantial gas reservoirs from infalling companions and thereby fuel elevated star-formation rates \citep{duan25, duan26_merger, puskas25_rate, puskas25}. 

We note that the standard EVS calculation used above does not condition on the observed large-scale environment or explicitly account for galaxy clustering. Extensions of the EVS framework show that conditioning on an overdense environment can shift the expected maximum stellar mass towards higher values, thereby reducing the apparent tension for galaxies residing in such regions \citep{jespersen25_merger}. S2 and S3 both lie in the broader overdense structure identified in GOODS-N at $z\simeq5.17$--$5.30$. However, applying the environment-conditioned framework requires a robust definition of the three-dimensional overdensity volume, its selection function and completeness, and its density relative to comparable subvolumes within the survey. These quantities are not sufficiently well established for the present sample to support a reliable quantitative calculation. Since S2 and S3 already lie within the $2\sigma$ interval of the unconditioned EVS calculation, accounting for their overdense environment would be expected to strengthen, rather than weaken, our conclusion that they are consistent with expectations from $\Lambda$CDM. A dedicated environment-dependent analysis is therefore left for future work.

Connecting S2 and S3 to lower-redshift descendants, we expect that these systems will quench their star formation on short timescales (within a few hundred Myr), since continued star formation would otherwise cause them to overshoot the observed high-mass end of the stellar mass function. This evolutionary pathway has been extensively discussed prior to \textit{JWST}, with numerous studies arguing that intensely star-forming DSFGs at $z\gtrsim4$ are natural progenitors of the massive quiescent galaxy population observed at $z\sim2-4$ \citep[e.g.,][]{toft14, simpson14, ikarashi15}, requiring short, highly efficient starburst phases with peak SFRs of several hundred $M_\odot\,\mathrm{yr}^{-1}$ followed by rapid quenching driven by gas exhaustion, feedback, and/or dynamical processes \citep{wellons15, tacchella16_MS, popping17_compact}. Such systems are preferentially found in overdense environments and proto-clusters, where mergers and enhanced gas accretion accelerate stellar mass assembly \citep{chiang13, casey16, overzier16, lim24}. Recent \textit{JWST} observations have now provided direct constraints on the stellar populations of massive quiescent galaxies at $z\approx3-4$ \citep{carnall23,glazebrook24,nanayakkara24,setton24,de-graaff25_qg,turner25,zhang26}, largely confirming this picture. For example, detailed \textit{JWST}/NIRSpec and NIRCam observations of ZF-UDS-7329, a massive quiescent galaxy at $z=3.2$, indicate a rapid formation history with a peak SFR of $400$--$800~M_\odot\,\mathrm{yr}^{-1}$ at $z\approx5$--$10$, consistent with a highly efficient, merger-driven starburst phase in an overdense environment \citep{turner25}. In this context, the high SFRs, disturbed morphologies, and dense large-scale environment of S2 and S3 are fully consistent with this long-standing DSFG-to-quiescent evolutionary pathway, now directly probed with \textit{JWST}. 

Our SED modelling cannot distinguish between stars formed in situ within the main progenitor and stars formed in other progenitors and subsequently accreted through mergers. The recovered SFHs should therefore be interpreted as the formation-time distributions of the stars currently associated with S2 and S3, rather than necessarily as the in-situ SFHs of their main progenitors. This distinction may be particularly relevant given their disturbed morphologies and the possible role of mergers in triggering their elevated SFRs \citep{cochrane25}. If a significant fraction of their stellar mass was assembled ex situ, less stellar mass would have formed within the main progenitor, lowering its inferred in-situ star-formation efficiency and further reducing any apparent tension with $\Lambda$CDM. We note, however, that ex-situ assembly does not change the present total stellar mass or an integrated efficiency defined as $M_\star/(f_{\rm b}M_{\rm halo})$ for the final system.

\subsection{Mass-to-light ratio and the dust attenuation law}
\label{subsec:ML_dust}

We now offer a more intuitive explanation for why the choice of dust attenuation law introduces significant uncertainty in estimating the mass-to-light ratio ($\mathrm{M_{\star}/L}$) and the stellar mass. The constraining photometry used in this work are the \textit{JWST} bands F182M, F210M and F444W. Those bands straddle the Balmer / 4000 \AA~ break at $z\approx5.5$, i.e., F182M and F210M lie short-ward of the 4000 \AA~ break, while F444W lies long-ward of it. Therefore, these bands provide crucial colour information that allows us to constrain the mass-to-light ratio ($\mathrm{M_{\star}/L}$). 

It is well known that observed colours and $\mathrm{M_{\star}/L}$ of stellar populations are correlated, and this behaviour can be used to infer stellar masses \citep[see, e.g.,][]{rudnick06, zibetti09}. In order to study the dependence of the $\mathrm{M_{\star}/L}$ on the F210M$-$F444W colour and stellar population parameters, we use FSPS \citep{conroy09a} to setup a grid in stellar age (SSP age), stellar metallicity (Z$_{\star}$), and dust attenuation. Specifically, we consider single stellar populations (SSPs) with stellar ages from 10 Myr to 1.5 Gyr, stellar metallicities from 0.1 to 1.0 Z$_{\odot}$, and dust attenuation values of A$_{\rm V}$ from 0 to 5. We explore the \citet{calzetti00} and SMC \citep{gordon03} dust curves. We fix the redshift to $z=5.5$ and assume no nebular emission.

Fig.~\ref{fig:ml_col} shows the resulting models of $\log\mathrm{M_{\star}/L_{\rm F444W}}$ as a function of the F210M$-$F444W colour. The left panel shows the stellar metallicity (different line styles) and dust attenuation (different colours) dependence, all with a fixed Calzetti dust attenuation law. In the right panel of Fig.~\ref{fig:ml_col}, we show how the $\mathrm{M_{\star}/L_{\rm F444W}}$ changes when changing the dust attenuation law from the Calzetti (solid line) to the SMC (dashed line), assuming solar metallicity for all lines. For all models, the SSP age increases along the lines as indicated by the arrow. As expected, the overall trend shows a higher $\mathrm{M_{\star}/L_{\rm F444W}}$ ratio for redder colours. At older ages (SSP age of $>100$ Myr), the variations in age, metallicity and amount of dust attenuation collapse onto a narrow relation between $\mathrm{M_{\star}/L_{\rm F444W}}$ and F210M$-$F444W colour. Hence, this strong degeneracy between stellar age, metallicity and amount of dust attenuation can be used to infer tight constraints on the $\mathrm{M_{\star}/L_{\rm F444W}}$ for a given F210M$-$F444W colour. At fixed colour, the dynamic range in $\mathrm{M_{\star}/L_{\rm F444W}}$ is roughly 0.3 dex at fixed attenuation curve. We stress that this relation between the M/L and the colour is not new. Similar arguments have been used on constrain the M/L of galaxies in the local Universe \citep{bell01, rudnick06, zibetti09} and at cosmic noon \citep{forster-schreiber11a, tacchella15_sci}. What we stress here is that the scatter in $\mathrm{M_{\star}/L_{\rm F444W}}$ significantly increases when variations in attenuation law are considered (see right panel of Fig.~\ref{fig:ml_col}), in particular for large attenuation values and red colours \citep[see also][]{markov23}. 

We use the measured fluxes (Tab.~\ref{tab:fluxes}) to infer the F210M$-$F444W colours of $2.21\pm0.11$, $4.02\pm1.68$ and $2.46\pm0.77$ AB mag for S1, S2 and S3, respectively. Using the above relation for the Calzetti law (assuming stellar ages of $>0.1$ Gyr), we infer a $\log\mathrm{M_{\star}/L_{\rm F444W}}$ in the range of $-0.27$ to $-0.13$, $0.42$ to $0.57$, and $-0.17$ to $-0.03$ for S1, S2, and S3, respectively. Assuming the SMC dust attenuation law, the inferred M/L ratios are lower: $-0.42$ to $-0.14$, $0.08$ to $0.37$, and $-0.35$ to $-0.07$ for S1, S2, and S3. Given the redshift of the sources, we can convert the F444W flux into a luminosity and obtain the stellar masses. For the Calzetti law and the above quoted M/L ratio, we obtain stellar masses in the range of $\log(M_{\star}/M_{\odot})=10.20-10.34$, $10.84-10.99$, and $10.06-10.20$ for S1, S2 and S3. Assuming the SMC law, the inferred stellar masses are in the range of $\log(M_{\star}/M_{\odot})=10.05-10.33$, $10.5-10.79$, and $9.88-10.17$. 

An additional systematic for the most FIR-luminous objects in our sample is the implication of extreme dust columns. For sources with $\sim$mJy-level dust continuum emission, the inferred dust masses and surface densities typically correspond to line-of--sight extinctions of $A_{\rm V}\sim100$ mag in compact DSFGs, as argued in pre-\textit{JWST} work based on submm-selected samples and, in particular, for ``NIR-faint'' systems \citep[e.g.,][]{simpson17_scuba, dudzeviciute20, smail21}. In this regime, rest-frame optical measurements do not fully sample the underlying stellar populations; instead, they preferentially trace the least-obscured sightlines, so that the observed optical-NIR SED can yield biased ages and mass-to-light ratios even before allowing for variations in the attenuation curve. Moreover, strong age-dependent obscuration (young stars embedded in dense birth clouds, older stars experiencing lower effective attenuation) is expected and observed in dusty starbursts, further complicating the mapping between observed rest-optical colours/breaks and the true, galaxy-integrated stellar mass \citep[e.g.,][]{poggianti00}. Consequently, at $z\gtrsim2$ the assumption that JWST/F444W provides a robust mass tracer can fail for the most obscured systems, in close analogy to earlier assumptions that HST/F160W was a reliable stellar-mass proxy at $z\gtrsim1$. 

Furthermore, in addition to the change in attenuation law discussed above, pushing to younger ages (see Fig.~\ref{fig:ml_col}) leads to a further drop in the mass-to-light ratio, thereby increasing the scatter in M$_\star$/L by about 0.4–0.5 dex at fixed colour. Furthermore, throughout this discussion, we have ignored nebular emission. As shown in the previous sections, we use \texttt{Prospector}, where we include nebular emission and a flexible attenuation law in order to robustly constrain the stellar masses of S1, S2, and S3; emission lines can significantly enhance the apparent strength of spectral breaks in broad-band photometry, since the filters trace both stellar continuum and line emission, and \texttt{Prospector} models this self-consistently while fitting for (and marginalising over) the attenuation law, thereby accounting for both effects simultaneously.

\section{Conclusions}
\label{sec:conclusions}

In this paper, we revisit the stellar mass estimates of three high-redshift, optically dark galaxies (S1, S2 and S3). \citet{xiao24} previously reported these galaxies to be among the most massive and actively star-forming systems at $z \approx 5$, inferring remarkably high star-formation efficiencies. Given their extreme properties, if confirmed, such galaxies could present a significant tension with predictions from galaxy formation models in the context of the $\Lambda$CDM cosmology. We expand upon previous work by including more observational data (\textit{JWST} imaging from $0.9-4.4~\mu$m, H$\alpha$ emission line flux, FIR data from ALMA/NOEMA) and perform a more flexible and self-consistent stellar population analysis. 

Recent \textit{JWST}/NIRSpec IFU observations securely identify multiple emission lines and yield a robust spectroscopic redshift of $z_{\rm spec}=3.2439 \pm 0.0002$ for S1 \citep{xiao26}. This revised redshift is significantly lower than the value of $z\approx5.6$ previously adopted by \citet{xiao24} and has important consequences for the inferred star-formation efficiency of S1. We then model the stellar populations of all three galaxies using \texttt{Prospector}, following the methodology described in \citet{tacchella22_highz, tacchella23_metal}, which enables a fully Bayesian inference of key physical properties such as stellar mass, SFH, and dust attenuation (see Section \ref{subsec:redshift} and Table \ref{table:priors and parameters}). For all three systems, the redshift is fixed to the spectroscopic value and the models are fit to the observed photometry and emission-line fluxes listed in Table~\ref{tab:fluxes}.

We show that for a fixed F210M$-$F444W colour, which straddles the Balmer / 4000 \AA~ break at $z\approx5.5$, stellar population models yield relatively tight constraints on the stellar mass-to-light ratio, with a dynamic range of $\sim0.3$ dex when assuming a fixed dust attenuation curve, consistent with relations established from the local Universe to cosmic noon. However, allowing for variations in the attenuation law substantially increases the uncertainty in $\mathrm{M_{\star}/L_{\rm F444W}}$ to $\approx1$ dex, particularly for heavily obscured and red systems.

Leveraging the flexibility of \texttt{Prospector}, we construct a fiducial stellar population model that adopts non-parametric SFHs with two different base priors: a rising SFH base and a constant SFH base. Our analysis reveals notable degeneracies between stellar mass, dust attenuation law, and the optical depth of diffuse dust, highlighting the sensitivity of stellar mass estimates to modelling assumptions. In general, the rising SFH prior yields lower stellar masses than the constant SFH prior, reflecting the reduced contribution from older stellar populations when recent star formation dominates the observed light.

Quantitatively (see Table~\ref{table:result}), the rising SFH base prior yields stellar masses that are lower by $\simeq0.2-0.4$ dex compared to the constant SFH base prior across all three galaxies (i.e., from $\log M_\star=10.36$ to $10.74$ for S1, $10.95$ to $11.22$ for S2, and $10.31$ to $10.51$ for S3), reflecting the systematically younger mass-weighted ages inferred under the rising SFH preference of the prior. In contrast, the inferred $\mathrm{SFR}_{50}$ values are relatively insensitive to the choice of SFH base prior, remaining consistent within uncertainties for all three sources despite the differences in the recovered stellar masses. Importantly, at these stellar masses and SFRs, all three systems are consistent with expectations from galaxy formation models in the context of the $\Lambda$CDM cosmology. However, the implied integrated star-formation efficiencies for S2 and S3 are still high, i.e., these systems formed very efficiently during the epoch of reionisation. 

Our analysis highlights the critical importance of carefully selecting the priors for SFH and dust attenuation when performing SED fitting, as they can substantially influence the inferred physical properties of high-redshift galaxies, in particular systems with high dust attenuation such as DSFGs. We also note that, for the most FIR-luminous objects, the implied dust columns can correspond to $A_{\rm V}\sim100$, meaning that rest-frame optical constraints (including F444W at $z\gtrsim2$) may not sample the full stellar population and should not be interpreted as a robust proxy for total stellar mass. In this high-extinction regime, both the uniqueness of complex SFH inferences from the optical–NIR SED and the interpretation of rest-optical morphologies warrant caution, and resolving these systematics will require deeper long-wavelength constraints and/or spatially resolved multi-band analyses.

\section*{Acknowledgements}

ST acknowledges support by the Royal Society Research Grant G125142. FDE acknowledges support by the Science and Technology Facilities Council (STFC), by the ERC through Advanced Grant 695671 ``QUENCH'', and by the UKRI Frontier Research grant RISEandFALL.
This work is based on observations made with the NASA/ESA/CSA \textit{JWST}. The data were obtained from the Mikulski Archive for Space Telescopes at the Space Telescope Science Institute, which is operated by the Association of Universities for Research in Astronomy, Inc., under NASA contract NAS 5-03127 for \textit{JWST}. These observations are associated with programs 1180, 1181, 1210, 1286, 1287, 1895, 1963, 3215, and 5572. The authors acknowledge the FRESCO team led by PI Pascal Oesch for developing their observing programme with a zero-exclusive-access period, and PI Mengyuan Xiao for developing the NIRSpec/IFU programme 5572 and providing the spectroscopic redshift. We also acknowledge the Hubble Legacy Fields team for the reduction and release of the HST ACS imaging we use in this work.

\section*{Data Availability}

The \textit{JWST} data used in this analysis is publicly available at MAST (\url{https://archive.stsci.edu/missions-and-data/jwst}). The best-fitting SED parameters for the galaxies in this analysis are available upon request.



\bibliographystyle{mnras}



\appendix

\begin{figure*}
    \centering
    \includegraphics[width=\linewidth]{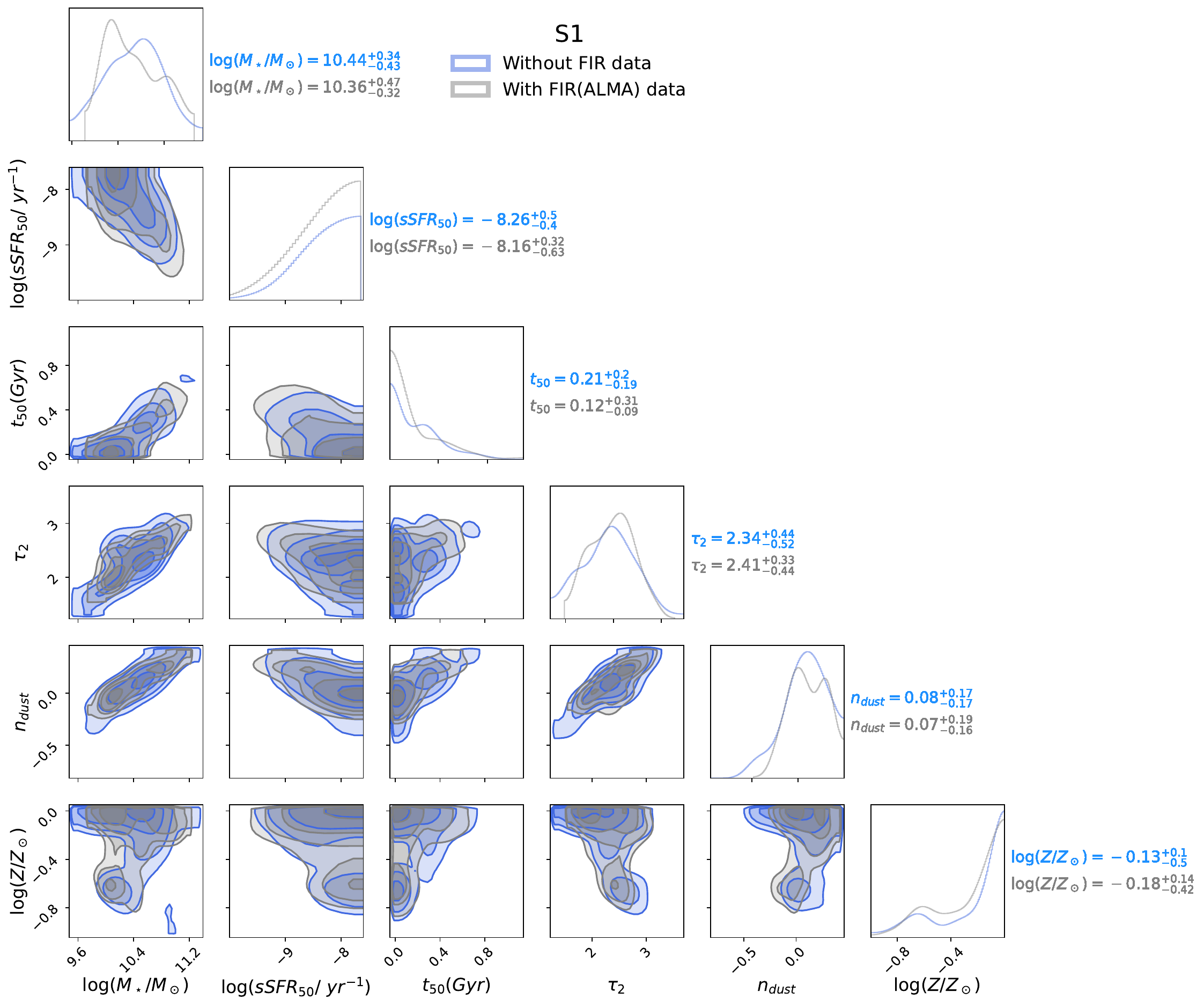}
    \caption{Effects of including FIR measurements (ALMA) in the SED fitting of galaxy S1. Posterior distributions for key parameters are shown for fits performed with (grey) and without (blue) the ALMA 1.1 mm data. The comparison highlights how the inclusion of FIR data influences the inferred physical properties.}
    \label{fig:corner_s1_noALMA}
\end{figure*}

\begin{figure*}
    \centering
    \includegraphics[width=\linewidth]{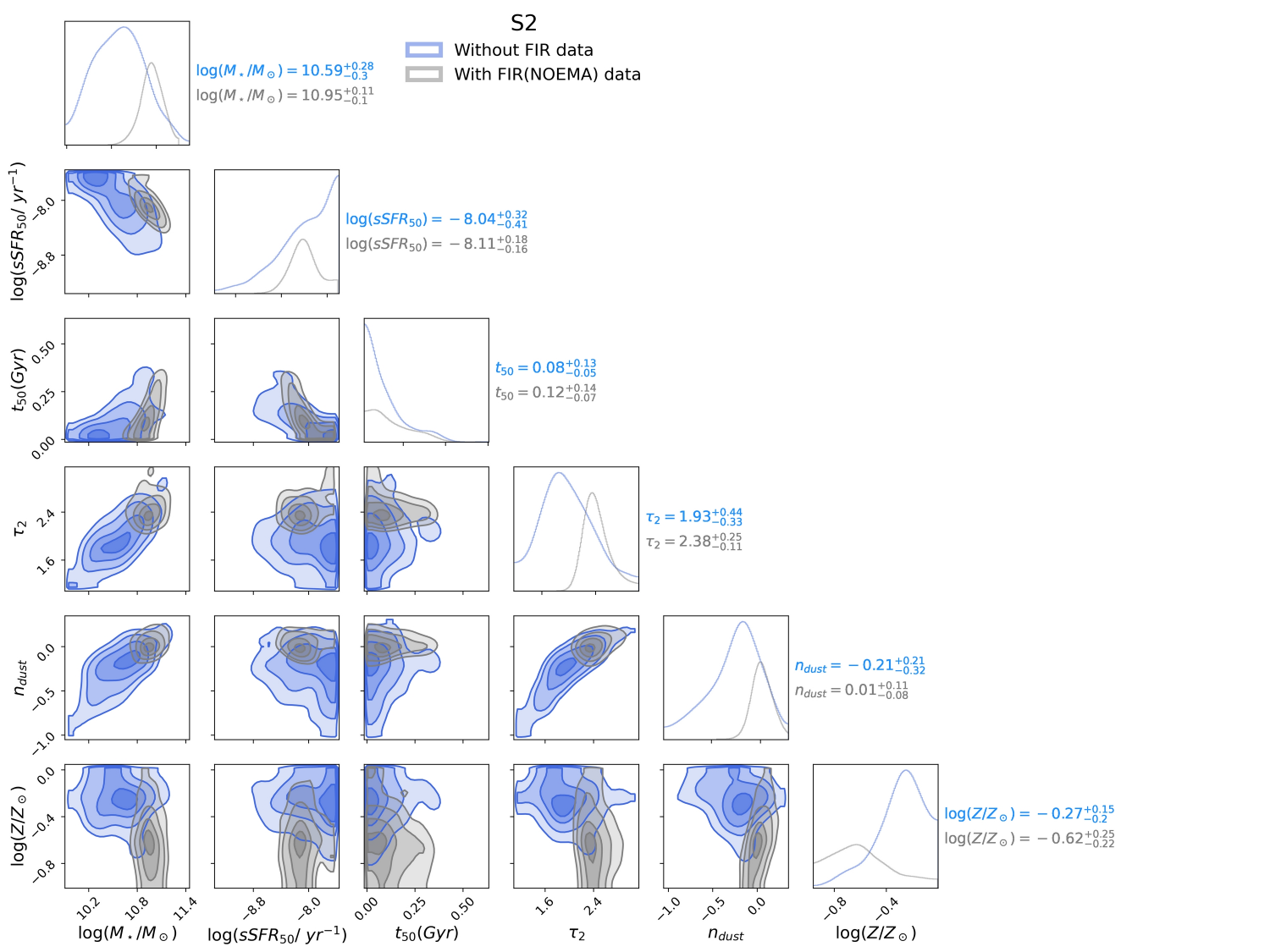}
    \caption{Same as Fig.~\ref{fig:corner_s1_noALMA} but for S2 with NOEMA measurements.}
    \label{fig:corner_s2_noNOEMA}
\end{figure*}

\section{Effect of including FIR constraints}
\label{app:stellar_mass_FIR}

To tightly constrain the posterior distributions of galaxy properties, our fiducial model fits (Section \ref{subsec:stellar_mass_dust_law}) incorporate all available photometry, including NIR fluxes from \textit{JWST}, FIR fluxes from ALMA and NOEMA for S1 and S2, respectively, and H$\alpha$ emission line measurements where available (see Table \ref{tab:fluxes}). To assess the impact of the FIR data, we perform additional fits excluding these constraints and compare the resulting posteriors. This allows us to isolate the influence of rest-frame FIR fluxes on parameters such as dust attenuation, star formation rates, and stellar mass, and to evaluate the robustness of the inferred properties in the absence of long wavelength data. \par

Figures~\ref{fig:corner_s1_noALMA} and \ref{fig:corner_s2_noNOEMA} present the posterior distributions for S1 and S2, respectively, comparing runs with (grey) and without (blue) FIR data under a rising SFH prior. For S1, the addition of the single ALMA 1.1 mm flux measurement produces only subtle improvements in the parameter constraints. The overall posterior structure, including the parameter degeneracies, remains largely unchanged, and the inferred stellar mass is nearly identical between the two runs. In contrast, S2 shows a much stronger response to the inclusion of the four NOEMA measurements. While the available H$\alpha$ measurement provides an additional constraint on the recent star formation rate, the broader FIR coverage also plays an important role. Unlike the single ALMA point, the multiple NOEMA measurements improve the SED fit, leading to tighter constraints on the dust properties and, through the energy-balance assumption, reducing degeneracies with the stellar population parameters. As a result, the posterior distributions for nearly all parameters become significantly narrower when the FIR data are included. \par

\section{Measurement of the Pa$\gamma$ flux for S1}
\label{app:pagamma}

We independently inspected the NIRSpec/IFU observations of S1 to verify the spectroscopic redshift and measure the Pa$\gamma$ flux used in Section~\ref{subsec:redshift}. We extracted a spatially integrated spectrum using an elliptical aperture with a semi-major axis of $0.3$ arcsec, ellipticity 0.315, and position angle $72.8^\circ$. The resulting spectrum contains emission lines at approximately 3.846, 4.042, 4.593, and $4.639~\mu$m, which we identify as [S{\sc iii}]$\lambda9069$, [S{\sc iii}]$\lambda9531$, He{\sc i}$\lambda1.083\mu$m, and Pa$\gamma$, respectively. These multiple line identifications independently confirm that S1 lies at $z\simeq3.24$, consistent with the spectroscopic redshift reported by \citet{xiao26} and adopted throughout this work.

The [S{\sc iii}]$\lambda9531$ and He{\sc i} lines exhibit double-peaked profiles. Motivated by the NIRCam imaging, in which a blue substructure (S1B) is visible to the south-west of the main source, we model the spectrum using two kinematic components associated with S1 and S1B. We simultaneously fit [S{\sc iii}]$\lambda\lambda9069,9531$, He{\sc i}$\lambda1.083\mu$m, and Pa$\gamma$. Within each component, the line centroids and intrinsic velocity dispersions are tied, while the [S{\sc iii}]$\lambda9531$/[S{\sc iii}]$\lambda9069$ flux ratio is fixed to 2.469 \citep{luridiana15}. The local continuum in each spectral window is described by a first-order polynomial, and the model is convolved with the NIRSpec line-spread function. Following an initial least-squares fit, we rescale the uncertainty vectors to match the scatter in the residuals, accounting approximately for correlated noise within the aperture. We then sample the posterior using \textsc{emcee} \citep{foreman-mackey13}. A two-component model is strongly preferred over a single component, with $\Delta{\rm BIC}=-35$.

For the primary S1 component, we measure $F_{\mathrm{Pa}\gamma} = (0.40\pm0.10)\times10^{-18}\,\mathrm{erg\,s^{-1}\,cm^{-2}}$, corresponding to an approximately $4\sigma$ detection. For S1B, we obtain $F_{\mathrm{Pa}\gamma,\mathrm{S1B}}  = (0.18\pm0.08)\times10^{-18}\,\mathrm{erg\,s^{-1}\,cm^{-2}}$. The total flux within the aperture is therefore $(0.58\pm0.13)\times10^{-18}~\mathrm{erg~s^{-1}~cm^{-2}}$. Adopting a Calzetti attenuation law with $A_V$=5.23, the Case B Pa$\gamma$-to-H$\alpha$ ratio of 0.031, and the \citet{kennicutt09} H$\alpha$–SFR calibration converted to a Chabrier IMF, the measured Pa$\gamma$ for S1 flux implies an SFR of approximately $49\pm12 \,M_\odot\,\mathrm{yr}^{-1}$. If instead we use the total Pa$\gamma$ flux within the aperture, including the contribution from S1B, we obtain an SFR of approximately $71\pm16\,M_\odot\,\mathrm{yr}^{-1}$.
A complete analysis of the NIRSpec/IFU observations, including the spatially resolved line emission and kinematics, is presented by \citet{xiao26}.

\begin{figure*}
\centering
\includegraphics[width=\textwidth]{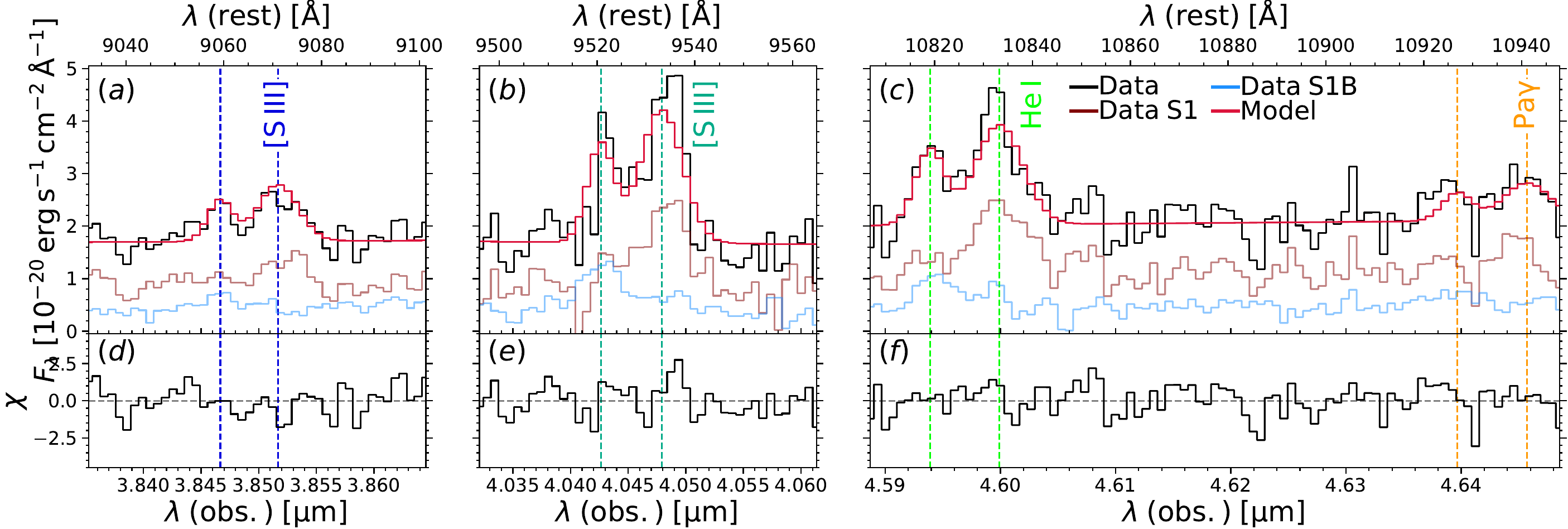}
\caption{NIRSpec/IFU spectrum of S1 in the spectral windows containing [S{\sc iii}]$\lambda9069$, [S{\sc iii}]$\lambda9531$, He{\sc i}$\lambda1.083\mu$m, and Pa$\gamma$. The black line shows the spectrum extracted within the elliptical aperture, and the red curve shows the maximum-likelihood two-component model. The separate spectra extracted around S1 and the south-western component S1B are also shown, illustrating the origin of the double-peaked line profiles. Residuals are displayed in the lower panels.}
\label{fig:s1_specfit}
\end{figure*}


\bsp	
\label{lastpage}
\end{document}